\documentclass[preprint,floatfix] {revtex4} 
\newcommand{\rvec}{\mathrm {\mathbf {r}}} 
\newcommand{\pvec}{\mathrm {\mathbf {p}}}

\usepackage{graphicx}
\usepackage{subfigure}
\usepackage{xcolor}
\usepackage{amsmath}
\usepackage{enumerate}
\usepackage{longtable}
\usepackage{tabularx}
\usepackage{multirow}
\usepackage{amssymb}
\usepackage{color, soul}
\definecolor{darkblue}{rgb}{0,0,0.5}
\setulcolor{darkblue}

\begin{document}

\title{Confined H$^-$ ion within a density functional framework}

\author{Sangita Majumdar}

\author{Neetik Mukherjee}

\author{Amlan K.~Roy}
\altaffiliation{Corresponding author. Email: akroy@iiserkol.ac.in, akroy6k@gmail.com.}
\affiliation{Department of Chemical Sciences\\
Indian Institute of Science Education and Research (IISER) Kolkata, 
Mohanpur-741246, Nadia, WB, India}

\begin{abstract}
Ground and excited states of a confined negative Hydrogen ion has been pursued under Kohn-Sham density functional approach by invoking a 
physically motivated work-function-based exchange potential. The exchange-only results are of near Hartree-Fock
quality. Local parameterised Wigner-type, and gradient- and Laplacian-dependent non-local Lee-Yang-Parr functionals are chosen to 
investigate the electron correlation effects. Eigenfunctions and eigenvalues are extracted by using a generalized pseudospectral method 
obeying Dirichlet boundary condition. Energy values are reported for 1s$^{2}$ ($^{1}$S), 1s2s ($^{3,1}$S) and 1s2p ($^{3,1}$P) 
states. Performance of the correlation functionals in the context of confinement is examined critically. The present results are in excellent 
agreement with available literature. Additionally, Shannon entropy and Onicescu energy are offered for ground and low lying singly excited 
1s2s ($^{3}$S) and 1s2p ($^{3}$P) states. The influence of electron correlation is more predominant in the weaker confinement limit 
and it decays with an increase in confinement strength. In essence, energy and some information measures are estimated using a
newly formulated density functional strategy.  



\end{abstract}
\maketitle

\section{Introduction}
Atomic and molecular systems confined by different forms of external potentials show various novel and interesting 
properties which are significantly different from their free counterparts. Although the study of confined atoms 
started several decades earlier \cite{michels37, sommerfeld38}, spectroscopic analysis of energy 
levels and other structural properties of quantum systems under diverse external confinements have received attention 
in recent years \cite{laughlin09, sabin09, flores-riveros10, yakar11, montgomery13, bhattacharyya13, sen14, 
montgomery15, saha16, galvez17, jiao17, chandra18}. An atom under spatial constraints may be modeled for describing 
the effect of pressure on the system, which may impact the rearrangement of orbitals, energy spectrum, continuum 
lowering; also, bonding pattern and co-ordination number may undergo dramatic changes in a 
molecule. Interested reader can find some elegant reviews in the literature \cite{jaskolski96, sabin09, sen14, leykoo18}. 
Such changes in structure play a crucial role for gaining insight to the unusual physico-chemical properties 
in constrained systems. A systematic analysis of one- and two-electron atom/ion is, therefore, essential for a 
comprehensive understanding of quantum confinement.  

A simple but interesting two-electron confined model is the hydrogen negative ion (H$^-$) restricted by a spherical barrier. 
Investigation on negative ions is an important research activity in atomic physics in their own right. Usually they are fragile quantum 
systems possessing binding energies less than one order of magnitude than that in the atom. H$^-$ ion, in particular, plays a fundamental 
role in the understanding of effect of correlation in three-body quantum mechanical problems. As a result of this weaker
binding, the correlation effects are rather sensitive and delicate, compared to an iso-electronic atom or positive 
ion. Several excellent reviews are available \cite{buckman94,miller17} on the subject. Almost eighty percent atoms are 
able to form stable negative ion. They play a dominant role in the context of 
electrical conductivity in weakly ionised gases and plasmas. The versatility of hydride ion 
has been well established. It acts as an efficient antioxidant in human body. In transition 
region of planetary nebula, it is present in high concentration. It also functions as the main source of opacity in 
sun atmosphere at red and infrared region. Unlike other negative ions, extensive theoretical study for H$^{-}$ ion has 
been done since 1962. However similar works on its \emph{confined} counterpart remains quite limited. 

Most of the studies in literature have considered He atom as the prototypical two-electron confined system. 
Spatially confined H$^-$ works are not so prevalent in literature, relatively speaking. Nevertheless, a 
decent number of methods exist. Some of these are: Hartree-Fock (HF) calculation with 
B-spline method \cite{wilson10}, a combination of quantum genetic algorithm and HF \cite{yakar11}, Hylleraas 
type wave function for variational calculation \cite{gimarc67}, quantum Monte Carlo \cite{joslin92}, CI calculation 
using explicitly correlated Hylleraas basis \cite{montgomery15}, for ground and singly excited S states. Apart from that, 
there prevails a couple of Rayleigh-Ritz approaches: (a) with three-parameter correlated wave function 
for ground state \cite{lesech11} (b) using explicitly correlated Hylleraas-type basis set for singly excited 1s2s and 
1s3s ($^1$S) \cite{saha16}, for 1s$^2$ ($^1$S), 2p$^2$, 1snp ($^{1,3}$P) with (n = 2--5) ($^3$P), in \cite{chandra18}. 
One also finds variational method based on (a) generalized Hylleraas basis (GHB) \cite{flores-riveros08, bhattacharyya13} and
(b) B-splines basis \cite{tong-yun01} as well. A detailed analysis of electron correlation has been published in 
\cite{wilson10}. A density functional theory (DFT) report is available in \cite{sen05}, within LDA and BLYP 
functional. Penetrable walls have also been undertaken as well. For example, energy spectrum for different confinement strengths
are analyzed for H$^-$ ion confined by an anisotropic harmonic oscillator potential, by (i) CI method within gaussian basis 
\cite{sako03} (ii) adiabatic hyper-spherical approach \cite{fang07}. Other than energy, properties like static dipole 
polarizability \cite{holka05, choluj17}, second hyperpolarizability \cite{choluj17} for are also pursued.
Energy levels and electric dipole polarizabilties of \emph{endohedrally} confined H$^-$ ion with CI method coupled with a 
B-spline approach are analyzed in \cite{melono18}. Some works are also reported in the context of H$^-$ 
ion embedded in plasma environment \cite{zhang96, winkler96, kar04, kar05, kar07, kar08, kar08pla, kar09, kar11, ho12, kar13, 
jiang13, jiao14, kar17}.

The relation between information theoretic tool and quantum mechanical kinetic energy was established in \cite{sear80}.
Since then the importance of these measures in the context of DFT has been discussed in several papers \cite{romera04,romera05,
nagy08,ghiringhelli10,nagy14a}. In a recent work the Euler equation in orbital-free DFT  is formulated by invoking 
Shannon entropy ($S$) and Fisher 
information \cite{nagy14}. Over the years these tools have emerged as versatile descriptors in analysing atoms and 
molecules \cite{guevara03,guevara05,moustakidis05,sen05}. They are functionals of density and can quantify it accurately 
in various complementary ways. In present work, 
we are specifically interested in two such measures, namely, Shannon entropy and Onicescu energy ($E$). The former is the arithmetic 
mean of uncertainty and can characterize a given density distribution in global way. The latter refers to the expectation value of density and 
generally complements the behavior of $S$. A decent amount of research work has been published to inspect these measures in \emph{free} 
atom/ion. However, in confined situation, parallel reports are quite limited and scattered. One can mention the works on  
confined H atom (CHA), where $S$ was studied with change in $r_c$, in composite $r,p$ spaces, in case of both 
$l = 0$ and \emph{non-zero} $l$ states \cite{sen05, jiao17, mukherjee18, mukherjee18a}. It was found that effect of 
confinement is more profound on higher states. Study of $S$ was also performed in \cite{sanchez19} for the hydrogen atom submitted to four 
different potentials: (a) infinite potential (b) Coulomb plus harmonic oscillator (c) constant potential and (d) dielectric continuum.
In many-electron atoms, $S$ has been explored mostly using correlated Hylleraas-type wave 
function, in either attractive or repulsive conditions. Some DFT works are also reported. Thus ground state-$S$ was 
considered for two-electron iso-electronic series (H$^{-}$ He, Li$^{+}$, Be$^{2+}$) under \emph{hard} (impenetrable rigid wall) confinement, 
by using the BLYP XC functional \cite{sen05}; another DFT study for ground and excited states is recently published in \cite{majumdar20} 
for He, Li$+$ and Be$^{2+}$. Of late, there is a growing interest to treat the so-called \emph{finite} (soft) confinement as 
well. Besides ground state, some limited works exist on low-lying excited states of $S$--mostly, for 
single \cite{ou17} and double \cite{ou17cpl} excitations in He.


Thus it appears that there is a need for DFT calculation for confined many electron systems, in particular the negative 
ions. The motivation of the present work lies in that. Here we perform a detailed and systematic study of 
energy as well as $S,E$, in composite $r$ and $p$ spaces, for ground and some low-lying singly excited states 
of H$^{-}$ ion, trapped inside high pressure environment. This is accomplished by invoking a simple work-function based 
exchange potential, motivated from physical grounds. The correlation effect is incorporated by using (i) a local, 
parameterized Wigner-type functional and (ii) the popular Lee-Yang-Parr (LYP) functional. The relevant KS differential 
equation under Dirichlet boundary condition is solved by adopting an accurate and efficient generalized pseudo-spectral 
(GPS) scheme. This procedure has been effectively applied to ground and a large number of excited states in free atoms 
as well as in some confinement works, with considerable success. Electron density, $S_{\rvec}, E_{\rvec}$ are estimated 
from self-consistent orbitals. The momentum-space orbitals are obtained by performing Fourier transformation to 
$r$-space orbitals in usual way. Electron momentum density is constructed from $p$-space orbitals and subsequently 
$S_{\pvec}, E_{\pvec}$ are obtained therefrom. Our pilot calculation are done on ground and 1s2s, 1s2p excited states. 
Section II sums up the adopted methodology. Section III imprints the calculated results along with a comparison with 
available references. Finally, Sec.~IV concludes with the outlook and future prospects. 
            
\section{Methodology} 
Here we briefly outline the proposed density functional method for a particular state of an arbitrary 
atom centered inside an impenetrable spherical cavity, followed by the GPS scheme for calculation of eigenvalues and 
energies of KS equation. This has been very successful for ground and various states (such as singly, 
doubly, triply excited states corresponding to low- and high-lying excitation, valence and core excitation, 
autoionizing, hollow, doubly hollow, Rydberg and satellite states etc.) of \emph{free or unconfined} 
neutral atoms as well as ions \cite{roy97, roy97a, roy97b, roy02, roy04, roy05, roy07}. Very
recently, this has been extended to confinement situations \cite{majumdar20}. Our focus remains on essential portions, 
omitting the relevant details, which are available in above references.  

Our starting point is the non-relativistic single-particle time-independent KS equation with imposed confinement, which 
can be conveniently written as (atomic unit employed unless otherwise mentioned), 
\begin{equation}
\left[ -\frac{1}{2} \nabla^2 +v_{eff} (\rvec) \right] \phi_i(\rvec) = \varepsilon_i \phi_i(\rvec),
\end{equation}
where the ``effective" potential is constituted of following terms, 
\begin{eqnarray}
   v_{eff}(\rvec)& = & v_{ne}(\rvec) +\int \frac{\rho(\rvec^{\prime})}{|\rvec - \rvec^{\prime}|}
 \mathrm{d}\rvec^{\prime}+\frac{\delta E_{xc}[\rho(\rvec)]}{\delta \rho(\rvec)} + v_{conf}(\rvec) .  
\end{eqnarray}
In this equation, the first three terms in right-hand site correspond to usual electron-nuclear attraction, 
classical Hartree repulsion and XC potentials respectively. The following perturbation accounts for the desired 
confinement ($r_c$ refers to the radius of spherical cage),
\begin{equation} v_{conf} (\rvec) = \begin{cases}
0,  \ \ \ \ \ \ \ r \leq r_{c}   \\
+\infty, \ \ \ \  r > r_{c}.  \\
 \end{cases} 
\end{equation}

Despite the remarkable progress and success in ground-state electronic structure and properties of atoms/molecules, 
in past five decades, excited state-DFT has faced difficulties and challenges. This is mainly due to 
lack of (i) an analogous Hohenberg-Kohn theorem and (ii) an accurate, proper XC functional for a general excited state. 
This work intends to employ an exchange potential \cite{sahni90, sahni92}, which is derived 
from physical grounds. Accordingly, one can interpret exchange energy as resulting from an interaction between  
an electron at $\rvec$ and its Fermi-Coulomb hole charge density $\rho_{x}(\rvec,\rvec^{\prime})$ at $\rvec^{\prime}$. 
Thus it is given by, 
\begin{equation}
 E_{x}[\rho(\rvec)] = \frac{1}{2}\int\int\frac{\rho(\rvec)\rho_{x}(\rvec,\rvec^{\prime})}{|\rvec - \rvec^{\prime}|}
\ \mathrm{d}\rvec \ \mathrm{d}\rvec^{\prime}. 
\end{equation}
The unique local exchange potential $v_{x} (\rvec)$ for a given state, can then be defined as the work done in 
bringing an electron to the point $\rvec$ against the electric field arising out of its Fermi-Coulomb hole density, 
leading to the following form, 
\begin{equation}
v_{x} (\rvec)  = -\int_{\infty}^{r} \mathcal{E}_{x}(\rvec) \mathrm{d}l , 
\end{equation}
where the electric field may be defined as, 
\begin{equation}
\mathcal{E}_{x}(\rvec) = \int\frac{\rho_{x}(\rvec,\rvec^{\prime})(\rvec - \rvec^{\prime})}
{|\rvec - \rvec^{\prime}|^{3}} \ \mathrm{d}{\rvec^{'}}.
\end{equation}
One can write the Fermi hole in terms of orbitals as,
\begin{equation}
\rho_{x}(\rvec,\rvec^{\prime})=-\frac{\left|\gamma(\rvec,\rvec^{\prime})\right|^2}{2 \rho(\rvec)},
\end{equation}
where $\left|\gamma(\rvec,\rvec^{\prime})\right|=\sum_{i} \phi_{i}^{*}(\rvec)\phi_{i}(\rvec^{\prime})$ is 
single-particle density matrix, while $\rho(\rvec)$ corresponds to electron density, expressed in terms of occupied 
orbitals ($n_{i}$ implies occupation number) as,
\begin{equation}
 \rho(\rvec) = \sum_{i=1}^{N} n_{i}|\phi_i(\rvec)|^2.  
\end{equation}
While $v_x(\rvec)$ thus defined above, can be accurately calculated, one needs to approximate the unknown correlation 
potential $v_c(\rvec)$ for practical calculations. For this purpose, we employ two correlation functionals, namely,  
a Wigner-type \cite{brual78} and LYP \cite{lee88}. They have been chosen on the basis of their success in the context 
of excited states, which are recorded in the references \cite{roy97,roy97a,roy97b,roy02,roy04,roy05,roy07}. This will  
give the opportunity to examine and calibrate the performance of these functionals in current situation.  

By taking the $v_{x}(\rvec)$ and $v_{c}(\rvec)$ as above, we proceed towards the solution of resulting KS equation 
following the Dirichlet boundary condition. This is done here by adopting an accurate and efficient GPS prescription, 
providing a non-uniform, optimal spatial discretization. It is a simple but effective method giving excellent results on 
numerous physically and chemically relevant problems, such as \emph{singular} and non-singular \cite{roy02,roy04, roy05, roy07,
roy04a, roy04b, roy05a, roy05b}, Coulomb, H\'ulthen, Yukawa, logarithmic, spiked oscillator, Hellmann potential, 
etc., along with its recent extension to quantum confinement \cite{sen06, roy15, roy16}. Since the details are 
well established and documented, we skip these here and refer the interested reader to the references above.

The numerical $p$-space wave function is obtained by Fourier transforming of $r$-space counterpart, as follows,
\begin{equation}
\begin{aligned}
\xi(\pvec) & = & \int \phi(\rvec) \ e^{i {\pvec}.{\rvec}} \mathrm{d}{\rvec}.
\end{aligned}
\end{equation}

Here, $\xi(\pvec)$ needs to be normalized. The normalized $r$- and $p$-space densities are then expressed in the forms as 
$\rho(\rvec) = \sum_{i=1}^{N} n_{i}|\phi_{i}(\rvec)|^2$ and $\Pi(\pvec) = \sum_{i=1}^{N} n_{i}|\xi_{i} (\pvec)|^2$ 
respectively, where $n_{i}$ indicates the occupation number of the $i$th orbital.

Next, $S_{\rvec}, ~S_{\pvec}$ and Shannon entropy sum $S_{t}$ are defined as given below, 
\begin{equation}
\begin{aligned} 
S_{\rvec} & =  -\int_{{\mathcal{R}}^3} \rho(\rvec) \ \ln [\rho(\rvec)] \ \mathrm{d} \rvec, \ \ \    
S_{\pvec}  =  -\int_{{\mathcal{R}}^3} \Pi(\pvec) \ \ln [\Pi(\pvec)] \ \mathrm{d} \pvec,  \\
S_{t} & = \left[S_{\rvec}+S_{\pvec}\right] \geq 3(1+\ln \pi), \ \ \ \ \ \text{in 3 dimension}. 
\end{aligned} 
\end{equation}
Here both $\rho(\rvec)$ and $\Pi(\pvec)$ are normalized to unity.

All the computations are done numerically. The convergence is ensured by carrying out calculations with respect to variation 
in grid parameters, such as total number of radial points and maximum range of grid. It is generally observed that convergence 
is achieved relatively easily in the lower $r_c$ region compared to the $r_c \rightarrow \infty$ limit. All the quantities given 
in following tables and plots have been checked for above convergence.

\begingroup      
\squeezetable
\begin{table}
\caption {\label{tab:table1}Ground-state energy of radially confined H$^-$ for different $r_c$. See text for details.}
\begin{ruledtabular} 
\begin{tabular}{ c|cc|ccc }
  $r_c$    &   X-only     &  Literature & XC-Wigner & XC-LYP  & Literature \\
\hline
  0.05  & 3885.9257&  3885.925658\footnotemark[1]  & 3885.7425 & 3887.3010 & 3885.922469\footnotemark[3], 3885.870899\footnotemark[4],\\
  0.1   & 955.9361 &    &  955.7694 & 956.7881  & \\
  0.4   & 53.8264 &    &  53.7168  &  54.0735  & 54.145\footnotemark[7], 54.3573\footnotemark[11]\\
  0.5   & 33.1632 &  33.18974\footnotemark[2]  &  33.0645  &  33.3253  & 33.1120\footnotemark[5], 33.435\footnotemark[7], 33.11307\footnotemark[13] \\
  0.7   & 15.5897 &  15.59244\footnotemark[2]  &  15.5072  &  15.7069  & 15.5400\footnotemark[5], 15.756\footnotemark[7], 15.54087\footnotemark[13] \\
  0.9   &  8.6106 &   8.61166\footnotemark[2]  &   8.5395  &   8.6853  & 8.5621\footnotemark[5], 8.7073\footnotemark[7], 8.56299\footnotemark[13] \\        
  1.0   &  6.6375 &  6.637526\footnotemark[1],6.64209\footnotemark[2]  & 6.5709  &  6.6969  & 6.633326\footnotemark[3], 6.589644\footnotemark[4], 6.5897\footnotemark[5],\\  
        &         &                                                    &         &          &  6.7133\footnotemark[7], 6.59047\footnotemark[13] \\   
  1.2   &  4.1341      &  4.13699\footnotemark[2]  &   4.0749  &   4.1706  & 4.0875\footnotemark[5], 4.1826\footnotemark[7], 4.1492\footnotemark[11], \\ 
        &              &                           &           &           & 4.08820\footnotemark[13]\\
  1.4   &  2.6808      &  2.68124\footnotemark[2]  &   2.6275  &   2.7011  & 2.6354\footnotemark[5], 2.7112\footnotemark[7], 2.63603\footnotemark[13] \\
  1.8   &  1.1764      &  1.17666\footnotemark[2]  &   1.1315  &   1.1758  & 1.1330\footnotemark[5], 1.1834\footnotemark[7], 1.13357\footnotemark[13] \\
  2.0   &  0.7665      &  0.76664\footnotemark[2]  &   0.7248  &   0.7591  & 0.7240\footnotemark[5], 0.7231\footnotemark[6], 0.7245\footnotemark[9], \\
        &              &                           &           &           & 0.7659\footnotemark[7], 0.7677\footnotemark[11], 0.72663\footnotemark[12] \\
  2.5   &  0.1799      &    &   0.1442  &   0.1616  & 0.1394\footnotemark[5],0.1388\footnotemark[6], 0.167\footnotemark[7]\\
  2.8   &  $-$0.0123   &    & $-$0.0454 & $-$0.0347 & $-$0.051936\footnotemark[8], $-$0.051936\footnotemark[14]\\
  3.0   &  $-$0.1040   &  $-$0.10408\footnotemark[2]  & $-$0.1357 & $-$0.1284 & $-$0.1431\footnotemark[5], $-$0.1435\footnotemark[6], $-$0.143084\footnotemark[8],\\
        &              &                              &           &           & $-$0.124\footnotemark[7], $-$0.1427\footnotemark[9], $-$0.13915\footnotemark[12]\\
        &              &                              &           &           & $-$0.14271\footnotemark[13], $-$0.143084\footnotemark[14] \\
  4.0   &  $-$0.3420   &  $-$0.34209\footnotemark[2]  & $-$0.3685 & $-$0.3714 & $-$0.3790\footnotemark[5], $-$0.3794\footnotemark[6], $-$0.379037\footnotemark[8],\\
        &              &                              &           &           & $-$0.369\footnotemark[7], $-$0.3786\footnotemark[9], $-$0.3295\footnotemark[11], \\
        &              &                              &           &           & $-$0.37464\footnotemark[12], $-$0.37875\footnotemark[13] \\
  5.0   &  $-$0.4258   &  $-$0.425815\footnotemark[1]  & $-$0.4493 & $-$0.4564 & $-$0.438594\footnotemark[3], $-$0.461974\footnotemark[4], $-$0.462073\footnotemark[8], \\
        &              &                               &           &           & $-$0.4620\footnotemark[5], $-$0.4623\footnotemark[6], $-$0.4617\footnotemark[9]\\ 
        &              &                               &           &           & $-$0.456\footnotemark[7], $-$0.462073\footnotemark[14] \\                 
  6.0   &  $-$0.4595   &  $-$0.45954\footnotemark[2]  & $-$0.4813 & $-$0.4902 & $-$0.4958\footnotemark[5], $-$0.4958\footnotemark[6], $-$0.495772\footnotemark[8],\\
        &              &                              &           &           & $-$0.492\footnotemark[7], $-$0.4956\footnotemark[9], $-$0.4406\footnotemark[11], \\ 
        &              &                              &           &           & $-$0.49166\footnotemark[12], $-$0.49558\footnotemark[13] \\
 10.0   &  $-$0.4861   &  $-$0.486150\footnotemark[1],0.48614\footnotemark[2]  & $-$0.5056 & $-$0.5157 & $-$0.509209\footnotemark[3], $-$0.524688\footnotemark[4], $-$0.524688\footnotemark[8],\\
        &              &                                                       &           &           & $-$0.5247\footnotemark[5],$-$0.5239\footnotemark[6], $-$0.5245\footnotemark[9],\\ 
        &              &                                                       &           &           & $-$0.523\footnotemark[7], $-$0.52455\footnotemark[13],  $-$0.524688\footnotemark[14]\\            
 $\infty$  &  $-$0.4879   &  $-$0.487930\footnotemark[1], $-$0.48793\footnotemark[10]   & $-$0.5070 & $-$0.5177 & $-$0.514489\footnotemark[3], $-$0.527748\footnotemark[4], \\  
           &              &                               &           &           & $-$0.5277\footnotemark[5], $-$0.5278\footnotemark[6], $-$0.528\footnotemark[7],\\
           &              &                               &           &           &  $-$0.52775\footnotemark[10], $-$0.52481\footnotemark[12], $-$0.527751\footnotemark[14]  \\     
\end{tabular}
\end{ruledtabular}
\begin{tabbing}
$^{\mathrm{a}}$Ref.~\cite{wilson10}. \hspace{25pt}  \=
$^{\mathrm{b}}$Ref.~\cite{yakar11}. \hspace{25pt}  \= 
$^{\mathrm{c}}$E$^{1}_{RL}$ result of Ref.~\cite{wilson10}. \hspace{25pt}  \=  
$^{\mathrm{d}}$E$^{2}$ result of Ref.~\cite{wilson10}. \hspace{25pt}  \=  
$^{\mathrm{e}}$Ref.~\cite{flores-riveros08}. \hspace{25pt}  \=  \\
$^{\mathrm{f}}$Ref.~\cite{joslin92}.   \hspace{25pt}  \= 
$^{\mathrm{g}}$Ref.~\cite{sen05}. \hspace{25pt}  \=  
$^{\mathrm{h}}$Ref.~\cite{chandra18}. \hspace{25pt}  \=  
$^{\mathrm{i}}$Ref.~\cite{ting-yun01}. \hspace{25pt}  \=  
$^{\mathrm{j}}$Ref.~\cite{gimarc67}. \hspace{25pt}  \=  
$^{\mathrm{k}}$Ref.~\cite{marin92}.   \\  
$^{\mathrm{l}}$Ref.~\cite{lesech11}. \hspace{25pt}  \=  
$^{\mathrm{m}}$Ref.~\cite{melono18}. \hspace{25pt}  \=  
$^{\mathrm{n}}$Ref.~\cite{bhattacharyya13}. \hspace{25pt}  \=  
\end{tabbing}
\end{table}
\endgroup
  
\section{Result and Discussion}
At the onset it is convenient to mention a few general comments about the conferred results for compressed H$^-$ ion. 
Non-relativistic energies will be reported for ground 
1s$^{2}$ $^{1}$S and low lying single excited 1s2s $^{3,1}$S, 1s2p $^{3,1}$P states. Results on $S,E$ in composite 
r- and p-spaces will be presented for 1s$^{2}$ $^{1}$S, 1s2s $^{3}$S and 1s2p $^{3}$P states. All results are in atomic 
units, unless stated otherwise. In order to organize the data in an appropriate manner, three sets of energies
are attempted, \emph{viz.,} (i) exchange-only (ii) involving Wigner correlation (iii) considering LYP correlation. Throughout the 
discussion, these are termed as X-only, XC-Wigner and XC-LYP. Ground-state energies for confined H$^{-}$ ion investigated with some 
interest. Consequently a healthy amount of literature is available and they are compared with the present calculation whenever 
feasible. However, for excited state such attempt is very uncommon and only a handful of results are available to collate. Furthermore,
investigation of $S$ and $E$ for confined H$^-$ ion is very scarce. Except \cite{sen05} no such record is available for comparison.   

\begingroup           
\squeezetable
\begin{table}
\caption {\label{tab:table2}Energies of 1s2s $^{3,1}$S states of radially confined H$^-$ for different $r_c$. See text for details.}
\centering
\begin{ruledtabular} 
\begin{tabular}{c | c c c c | c c  c c}
 & \multicolumn{4}{c|}{$^{3}$S}   &  \multicolumn{4}{c}{$^{1}$S}  \\
\cline{2-5} \cline{6-9} 
$r_c$ & X-only & XC-Wigner & XC-LYP & Literature & X-only & XC-Wigner & XC-LYP & Literature \\
\hline
0.1  & 2426.7390 & 2426.5730  & 2428.1661 &  -                                 & 2432.4082 & 2432.2422 & 2433.8353  &  -      \\
0.2  &  596.4463 &  596.3056  &  597.2946 & -                                  & 599.3149  &  599.1741 &  600.1631  &   -     \\
0.5  &   90.4448 &   90.3468  &   90.8032 & 91.20906$^{a}$                     & 91.6346   & 91.5368   & 91.9931    &  -      \\
0.6  &   61.6373 &   61.5482  &   61.9275 & 62.15161$^{a}$                     & 62.6411   &  62.5522  & 62.9314    &  -      \\
0.9  &   25.8082 &   25.7377  &   25.9755 & 25.97643$^{a}$                     & 26.5030   &  26.4327  & 26.6703    &  -      \\
1    &   20.4697 &   20.4037  &   20.6111 & 20.52687$^{a}$, 20.4597$^{b}$      & 21.1030   & 21.0372   &  21.2444   &   -     \\
1.2  &   13.6038 &   13.5451  &   13.7057 & 13.64636$^{a}$, 13.5938$^{b}$      & 14.1451   & 14.0868   & 14.2470    &  -      \\
1.4  &    9.5381 &    9.4853  &    9.6118 &  9.56340$^{a}$,  9.5284$^{b}$      & 10.0142   & 9.9618    & 10.0879    &  -      \\
1.5  &    8.1075 &    8.0571  &    8.1699 &  -                                 & 8.5576    &  8.5077   & 8.6201     &  -    \\
1.8  &    5.2038 &    5.1594  &    5.2406 &  5.22940$^{a}$,  5.1946$^{b}$      & 5.5936    & 5.5499    & 5.6305     &  -      \\
2    &    3.9798 &    3.9387  &    4.0043 &  3.99015$^{a}$,  3.9709$^{b}$      & 4.3398    & 4.2993    & 4.3644     &  -      \\
3    &    1.2171 &    1.1864  &    1.2093 &  1.22002$^{a}$,  1.2095$^{b}$      & 1.4879    & 1.4579    & 1.4802     &  -      \\
4    &    0.3435 &    0.3184  &    0.3245 &  0.34429$^{a}$,  0.3371$^{b}$      & 0.5660    & 0.5410    & 0.5468     & -       \\
5    & $-$0.02131  & $-$0.04307 & $-$0.04430 & -                               & 0.1626    & 0.1396    & 0.1389     &  -      \\
6    & $-$0.2004  & $-$0.22004   & $-$0.2243 & $-$0.20046$^{a}$, $-$0.2050$^{b}$  & $-$0.0537 & $-$0.0759 & $-$0.0781  &  -      \\
8    & $-$0.3567  & $-$0.3738   &  $-$0.3772  & -                              & $-$0.2743 & $-$0.2945 & $-$0.2873  & -       \\
10   & $-$0.4181  & $-$0.4338   &  $-$0.4307  &  -               &    $-$0.3737          & $-$0.3913      &  $-$0.3682    & -                        \\
15   & $-$0.4689  & $-$0.4831   & -   &  -               &    $-$0.4562          & $-$0.4706             & -     &  -                       \\
\end{tabular}
\end{ruledtabular}
\begin{tabbing}
$^{\mathrm{a}}$Ref.~\cite{yakar11} (X-only energies). \hspace{25pt}    \=
$^{\mathrm{b}}$Ref.~\cite{flores-riveros08} (Correlated energies). \hspace{25pt}  \=  
\end{tabbing}
\end{table}
\endgroup     
 
\subsection{Energy analysis}
Let us begin the discussion with ground state energies of confined H$^{-}$ ion given in Table~\ref{tab:table1} at certain
representative $r_{c}$'s, starting from very strong confinement regime $(r_{c}=0.05)$ to free limit 
$(r_{c} \rightarrow \infty)$. Present X-only results are reported in second column. These outcomes are almost identical with 
HF results obtained by using B-spline approach employing zeroth order spherical Bessel function \cite{wilson10}. In this context, 
note that, an analogous agreement with HF calculation \cite{ludena78} is also observed in case of He-isoelectronic series and 
Li, Be atoms for which similar calculation has been done by the authors and it will be published soon \cite{majumdar20a}. Apart from 
that, the X-only values are also 
compared by invoking a combined quantum genetic algorithm (QGA) and RHF method \cite{yakar11}. A slightly compromised 
matching is observed at $r_{c} \le 1$ region. However similarity between these two results improves with rise in $r_{c}$. At moderate 
to large $r_{c}$ ($\ge 3$) both the results become identical. All these literature values are available in third column of 
Table~\ref{tab:table1}.

\begingroup           
\squeezetable
\begin{table}
\caption {\label{tab:table3}Energies of 1s2p $^{1,3}$P states of radially confined H$^-$ for different $r_c$. See text for details.}
\centering
\begin{ruledtabular} 
\begin{tabular}{c | c c c c | c c  c c}
 & \multicolumn{4}{c|}{$^{3}$P}   &  \multicolumn{4}{c}{$^{1}$P}  \\
\cline{2-5} \cline{6-9} 
$r_c$ & X-only & XC-Wigner & XC-LYP & Literature & X-only & XC-Wigner & XC-LYP & Literature \\
\hline
0.1  & 1472.6466  & 1472.4815  & 1473.7385  & -                & 1479.8278    & 1479.6627    & 1480.9194  & -       \\
0.2  &  360.5257  &  360.3864  &  361.1685  & -                &  364.1099   & 363.9706    & 364.7524     & -        \\
0.5  &   53.9750  &   53.8797  &   54.2403  & 54.02876$^{a}$   &   55.4003  & 55.3050    & 55.6653     &  -      \\
0.6  &   36.6134  &   36.5270  &   36.8264  & 36.64832$^{a}$   &   37.7986  & 37.7123    & 38.0113      & -       \\
0.9  &   15.0984  &   15.0309  &   15.2173  & 15.10315$^{a}$   &   15.8830  & 21.3313    & 16.0016     &  -      \\
1.0  &   11.9084  &   11.8453  &   12.0075  & 11.92050$^{a}$   &   12.6126  & 16.6321    & 12.7114     &  -      \\
1.2  &    7.8186  &    7.7629  &    7.8875  &  7.82685$^{a}$   &    8.4022 & 10.7516    &  8.4708    &   -     \\
1.4  &    5.4079  &    5.3580  &    5.4551  &  5.41271$^{a}$   &    5.9051 & 7.3797    &  5.9521    &   -     \\
1.5  &    4.5627  &    4.5152  &    4.6013  &  -               &    5.0252    & 6.2176    & 5.0636     &  -      \\
1.8  &    2.8547  &    2.8133  &    2.8737  &  2.85748$^{a}$   &    3.2360 & 3.9069    & 3.2547     &   -     \\
2    &    2.1393  &    2.10104 &    2.1488  &  2.14134$^{a}$   &    2.4796 & 2.9556    & 2.4889     &   -     \\
3    &    0.5434  &    0.5153  &    0.5281  &  0.54437$^{a}$   &    0.7585 & 0.8711    & 0.7430     &   -     \\
4    &    0.04955 &    0.02665 &    2.59005 &  0.04969$^{a}$   &    0.1981 & 0.2033    & 0.1743     &   -     \\
5    & $-$0.1555  & $-$0.17539 & $-$0.1817  & $-$0.162030$^{b}$ & $-$0.0497    & $-$0.0566    & $-$0.0757     &  $-$0.087731$^{b}$      \\
6    & $-$0.2586  & $-$0.27669 & $-$0.2850  & $-$0.25855$^{a}$, $-$0.264743$^{b}$    & $-$0.1832   & $-$0.1943    & $-$0.2083     &  $-$0.215790$^{b}$      \\
8    & $-$0.3575  & $-$0.3737  & $-$0.3803  & $-$0.362587$^{b}$    & $-$0.3199    & $-$0.3332    & $-$0.3364     &  $-$0.341509$^{b}$      \\
10   & $-$0.3957  & $-$0.4109  & $-$0.4063  &     -                 & $-$0.3951   & $-$0.4102    & $-$0.4041     &       -                  \\
15   & $-$0.4533  & $-$0.4671  & -          &     -                 & $-$0.4531   & $-$0.4669    & -             &       -                  \\
\end{tabular}
\end{ruledtabular}
\begin{tabbing}
$^{\mathrm{a}}$Ref.~\cite{yakar11} (X-only energies). \hspace{25pt}  \=
$^{\mathrm{b}}$Ref.~\cite{chandra18} (Correlated energies). \hspace{25pt}  \=  
\end{tabbing}
\end{table}
\endgroup 

The columns 4 and 5 of Table~\ref{tab:table1} now represent the Wigner and LYP energies respectively; with corresponding 
references in column 6. At strong confinement zone $(\approx r_{c} \le 3)$, Wigner energies are lower compared to LYP.
However, at moderate to large $r_{c}$ region an opposite behavior is seen. The difference between these two energies remains in the 
range of $-$0.0107 to 1.5585 Moving from free to confinement condition total energy increases. This happens mainly
due to an abrupt rise in kinetic energy. In most cases (except $r_c=0.05, 2.5, 2.8, 3$), either of the correlated present 
results (PR) shows appreciable agreement with explicitly correlated GHB \cite{wilson10,flores-riveros08,chandra18,gimarc67,bhattacharyya13}
energies. In XC-Wigner and XC-LYP, the absolute deviations are 0.003$\%$-3.93$\%$ and 0.03$\%$-4.84$\%$ respectively. At $r_{c}=0.05$, 
Wigner energies show excellent agreement with 
reported results, but a slight deviation is seen relative to XC-LYP value. However, at $r_{c}=2.5, 2.8, 3$, PR diverge from literature.
It is important to mention that, in almost all the cases XC-LYP values are higher than the best-possible results \cite{flores-riveros08,
chandra18,gimarc67,bhattacharyya13}, but no such trend is seen in XC-Wigner. These references also suggest that, at 
small $r_{c}$ region, Wigner performs better than LYP, but the scenario reverses with weakening of confinement strength.  
At strong confinement region $(r_{c}<4)$ PR are smaller than BLYP energies given in \cite{sen05}. However, this pattern 
reverses at $r_{c}>4$ range. Interestingly, at $r_{c}=4$,
Wigner and BLYP energies \cite{sen05} become almost identical. A similar situation arises for LYP functional at $r_{c}=5$. In essence 
both Wigner and LYP produce reasonably good agreement with the BLYP results \cite{sen05}, recording absolute deviations of 
0.13$\%$-13.65$\%$ and 0.08$\%$-3.54$\%$ respectively. At certain $r_{c}$'s ($\ge 0.4$) these are also tallied with CI method 
coupled with a B-Spline approach \cite{melono18}. In this case, the absolute deviation involving Wigner and LYP are 
0.14-4.91$\%$ and 0.64-10.02$\%$ successively. Besides these, PR produces good agreement with other correlated energies available in 
\cite{ting-yun01,marin92,lesech11}. It is needless to mention that, as usual both X-only and correlated energies abate with rise in $r_{c}$.

\begin{figure}                         
\begin{minipage}[c]{0.5\textwidth}\centering
\includegraphics[scale=0.90]{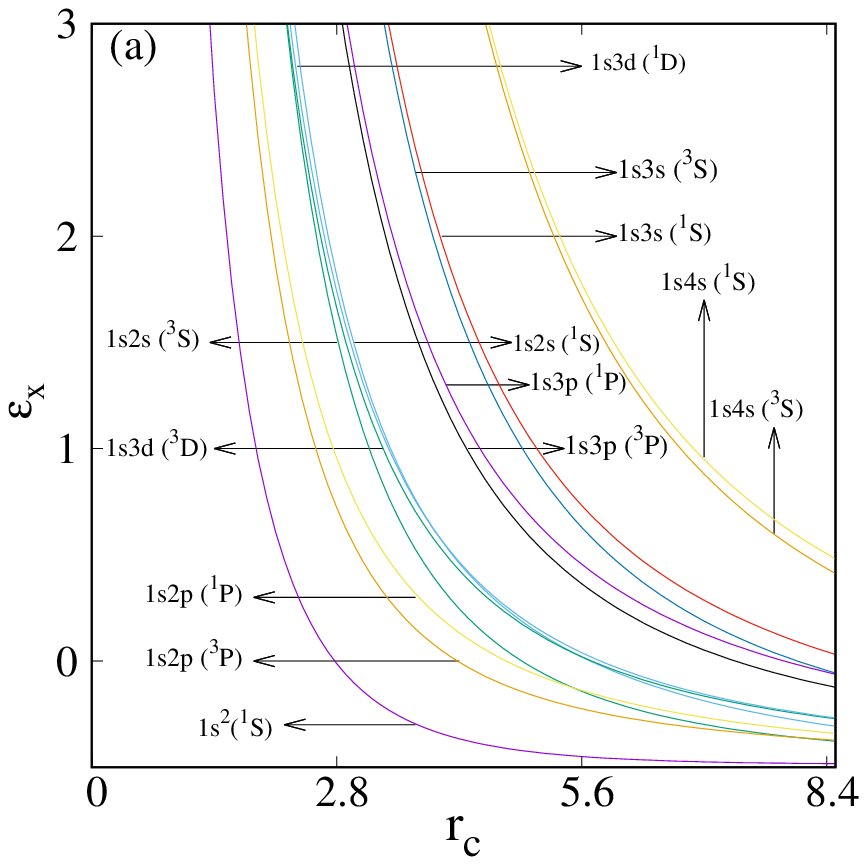}
\end{minipage}%
\begin{minipage}[c]{0.5\textwidth}\centering
\includegraphics[scale=0.90]{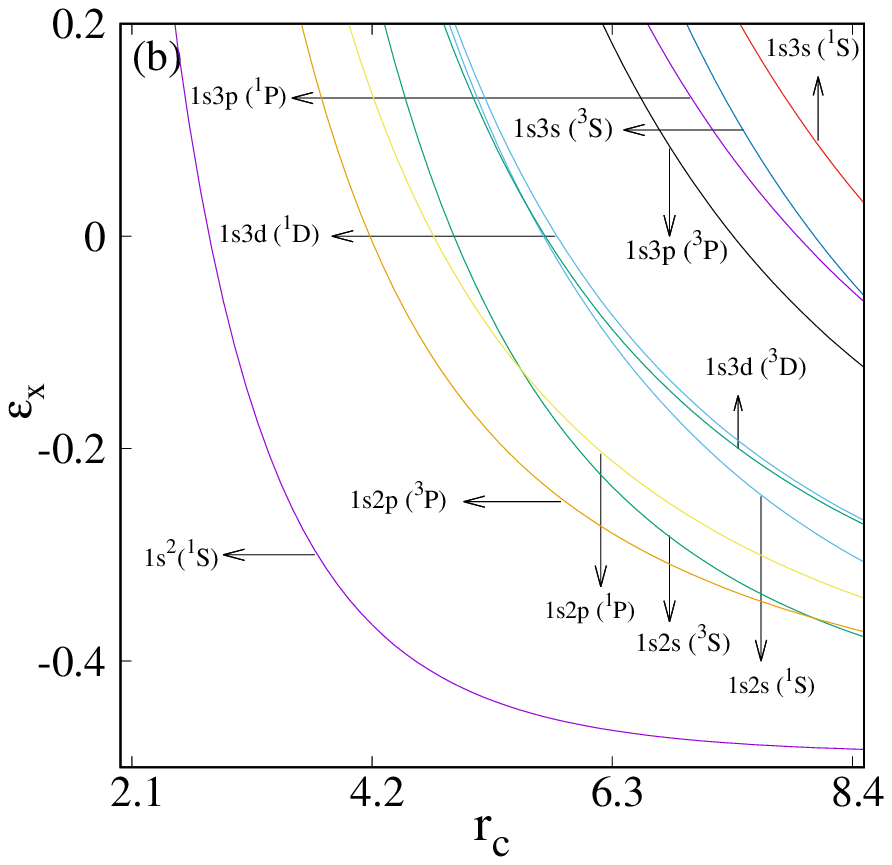}
\end{minipage}%
\caption{\label{fig:figure1}Energy changes in some low-lying states of confined H$^{-}$ with $r_c$ in (a). Panel (b) shows a 
magnification of (a) in $\epsilon \leq 0.2$ region. See text for details.}
\end{figure}   

Next, we move to employ this method in excited states. This provides an idea about its utility as well as performance 
in such states under hard confinement. Table~\ref{tab:table2} imprints energies of singly excited 1s2s 
$^{3}$S and $^{1}$S states of a trapped H$^{-}$ ion for a wide range of $r_c$. Reference theoretical results in this
context, are very rare. To the best of our knowledge, both X-only and correlated results are available only for $^{3}$S 
state and no such values are reported for $^{1}$S state. The second and sixth columns represent X-only 
results for triplet and singlet states. The X-only energies for $^{3}$S state can be compared with combined
QGA-RHF method \cite{yakar11} and the results show good agreement. Similar to the ground state, here also the 
convergence between these two values increases with progress in $r_{c}$. Third and fourth columns of Table~\ref{tab:table2} 
offer the wigner and LYP energies for $^{3}$S, whereas columns six and seven provide the same for $^{1}$S. 
Like the ground state, here also for both triplet and singlet states, LYP values are higher than Wigner data 
in $r_{c} \le 5$ region. Further, in $^{3}$S,
the matching between Wigner and LYP energies enhances with advancement in $r_{c}$. Absolute difference between 
these two correlated values for both the states are almost identical and it is in the range 0.003 to 
1.593. Present correlated energies for triplet state are in good agreement with the available GHB results \cite{flores-riveros08}. 
In this case, absolute deviation for Wigner and LYP are 0.27-7.33$\%$ and 0.016-9.41$\%$ successively. It may also
be noted that, in both cases deviation is higher in large $r_{c}$ regime. 

\begingroup           
\squeezetable
\begin{table}
\caption {\label{tab:table4}Energy values of some singly excited singlet and triplet states in radially confined H$^-$ at $r_c=0.1$. Energies
are arranged in ascending order. See text for details.}
\centering
\begin{ruledtabular} 
\begin{tabular}{c | c | c c c }
Configuration &States &  X-only & XC-Wigner & XC-LYP  \\
\hline
1s2p  & $^{3}$P, $^{1}$P  &  1472.6466, 1479.8278 & 1472.4815, 1479.6627  &  1473.7385, 1480.9194  \\
1s3d  & $^{3}$D, $^{1}$D  &  2127.2860, 2132.8279 & 2127.1219, 2129.8212  &  2128.6253, 2130.1494  \\
1s2s  & $^{3}$S, $^{1}$S  &  2426.7390, 2432.4082 & 2426.5730, 2432.2422  &  2428.2661, 2433.8353  \\
1s4f  & $^{3}$F, $^{1}$F  &  2909.1492, 2910.6018 & 2908.9858, 2910.4384  &  2910.7391, 2912.1918  \\
1s3p  & $^{3}$P, $^{1}$P  &  3445.2211, 3448.2732 & 3445.0558, 3448.1079  &  3446.9466, 3449.9986  \\ 
1s5g  & $^{3}$G, $^{1}$G  &  3815.8480, 3816.6892 & 3815.6851, 3816.5263  &  3817.6904, 3818.5316  \\
1s4d  & $^{3}$D, $^{1}$D  &  4600.3498, 4602.1893 & 4600.1849, 4602.0245  &  4602.3665, 4604.2060  \\
1s6h  & $^{3}$H, $^{1}$H  &  4844.9833, 4845.5142 & 4844.8208, 4845.3517  &  4847.0792, 4847.6102  \\
1s3s  & $^{3}$S, $^{1}$S  &  4891.8255, 4894.0532 & 4891.6597, 4893.7214  &  4893.9043, 4896.1319  \\
1s5f  & $^{3}$F, $^{1}$F  &  5891.7768, 5892.9728 & 5891.6124, 5892.6439  &  5894.0811, 5895.2772  \\
1s7i  & $^{3}$I, $^{1}$I  &  5994.5782, 5994.9344 & 5994.4160, 5994.7722  &  5996.9280, 5997.2843  \\
1s4p  & $^{3}$P, $^{1}$P  &  6403.9567, 6405.4891 & 6403.7913, 6405.3236  &  6406.3592, 6407.8914  \\
1s8k  & $^{3}$K, $^{1}$K  &  7263.0203, 7263.2706 & 7262.8583, 7363.1086  &  7265.6245, 7265.8746  \\
1s6g  & $^{3}$G, $^{1}$G  &  7317.1025, 7317.9240 & 7316.9384, 7317.7598  &  7319.6921, 7320.5136  \\
1s5d  & $^{3}$D, $^{1}$D  &  8055.1773, 8056.2696 & 8055.0122, 8056.1044  &  8057.8952, 8058.9874  \\
1s4s  & $^{3}$S, $^{1}$S  &  8343.8330, 8345.0383 & 8343.6671, 8344.8724  &  8346.6009, 8347.8063  \\
\end{tabular}
\end{ruledtabular}
\end{table}
\endgroup                        

After the successful attempt of the present method in 1s2s configuration we now arrive at 1s2p case to investigate its
$^{3}$P and $^{1}$P states under compression. Table~\ref{tab:table3} provides energies for these 
two states at same range of $r_{c}$ given in previous table. Similar to the earlier excited states, only a 
handful of literature is available and they are mentioned in the footnotes. As usual the X-only results for triplet and 
singlet states are given in columns two and six respectively. Again the $^{3}$P state corroborates with the QGA-RHF 
energies \cite{yakar11}. Moreover, akin to the previous two cases, the extent of convergence (with literature energies) 
promotes with growth in $r_{c}$. Here
also no literature is available to check the $^{1}$P results. Correlated energies for $^{3}$P and $^{1}$P 
are presented in columns 3,~4 and 7,~8. The absolute difference between the two correlated
results are 0.0046-1.257 and 0.0032-1.2567 respectively. Moreover, both 
values approach each other with relaxation in confinement. In either of the cases GHB results are available for 
$r_{c}=5, 6, 8$ \cite{chandra18}, which offer reasonable agreement with PR.  

\begin{figure}                         
\begin{minipage}[c]{0.50\textwidth}\centering
\includegraphics[scale=0.8]{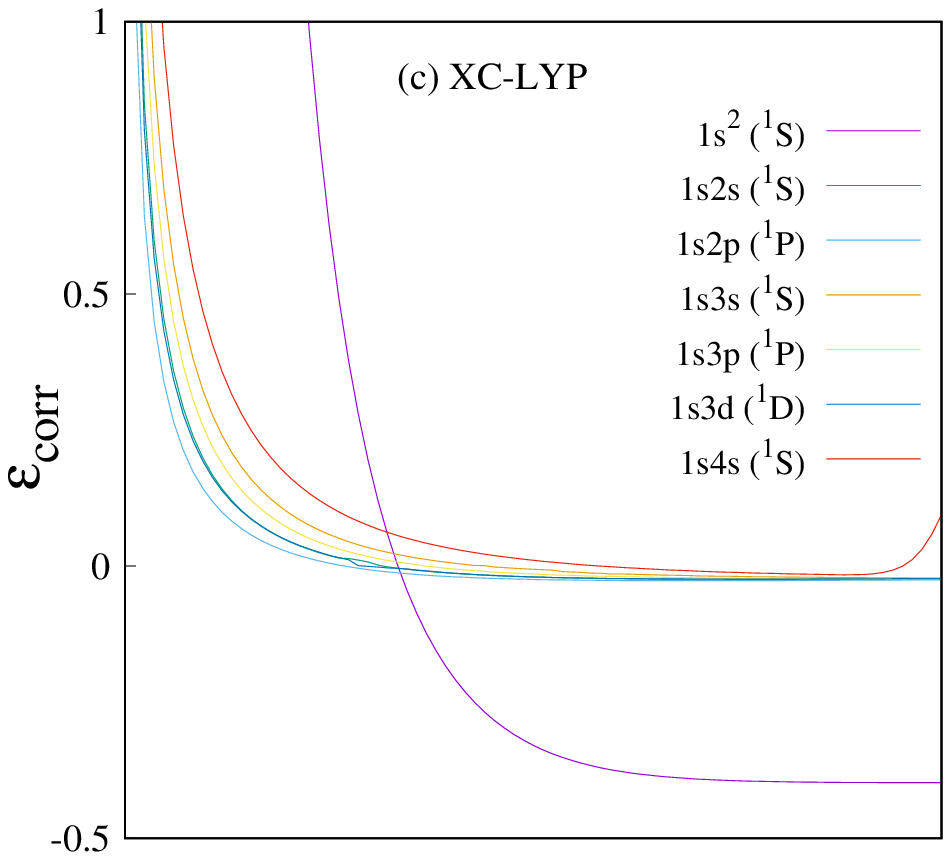}
\end{minipage}%
\begin{minipage}[c]{0.50\textwidth}\centering
\includegraphics[scale=0.8]{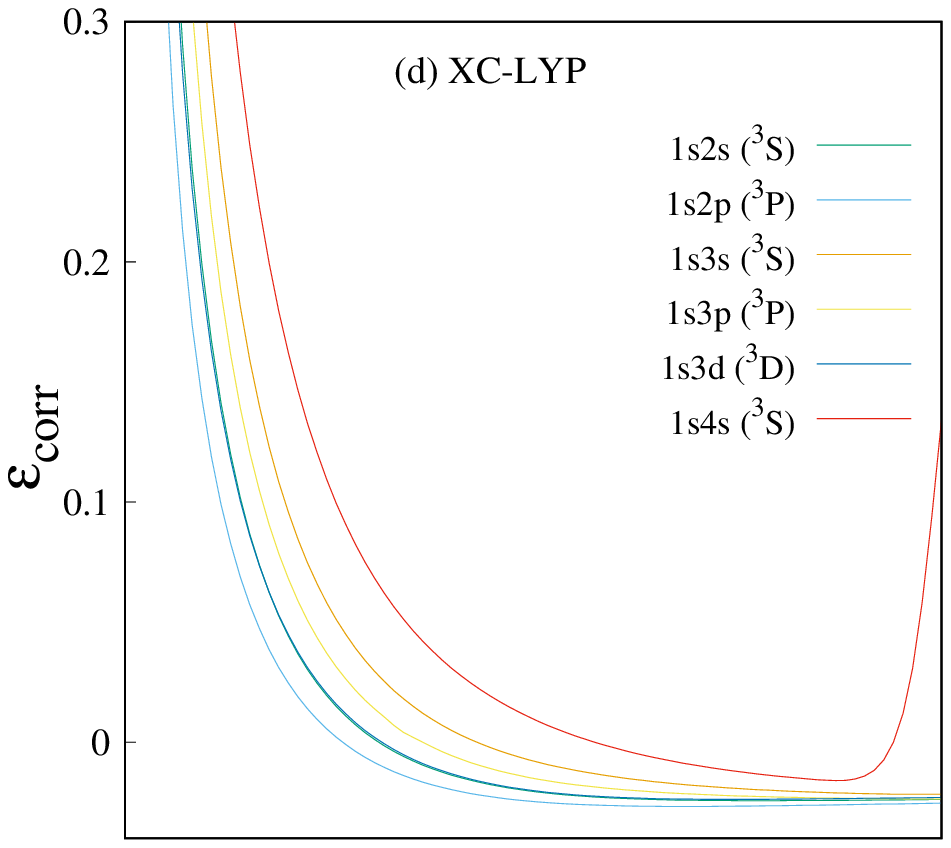}
\end{minipage}%
\vspace{5mm}
\hspace{0.2in}
\begin{minipage}[c]{0.50\textwidth}\centering
\includegraphics[scale=0.85]{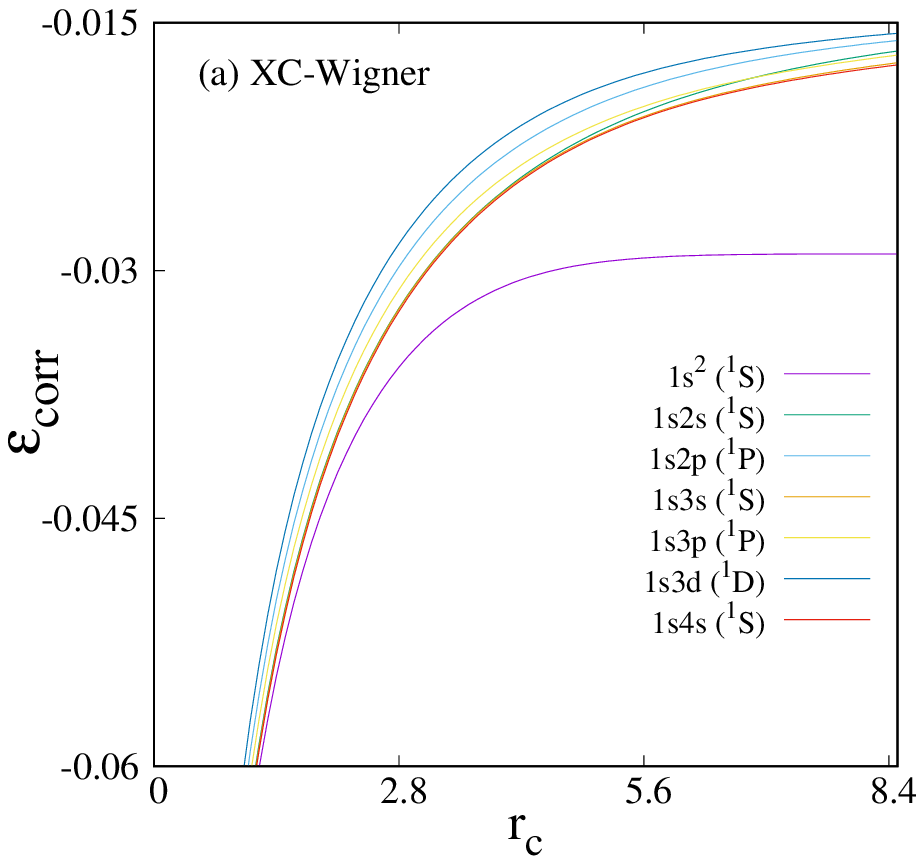}
\end{minipage}%
\begin{minipage}[c]{0.50\textwidth}\centering
\includegraphics[scale=0.85]{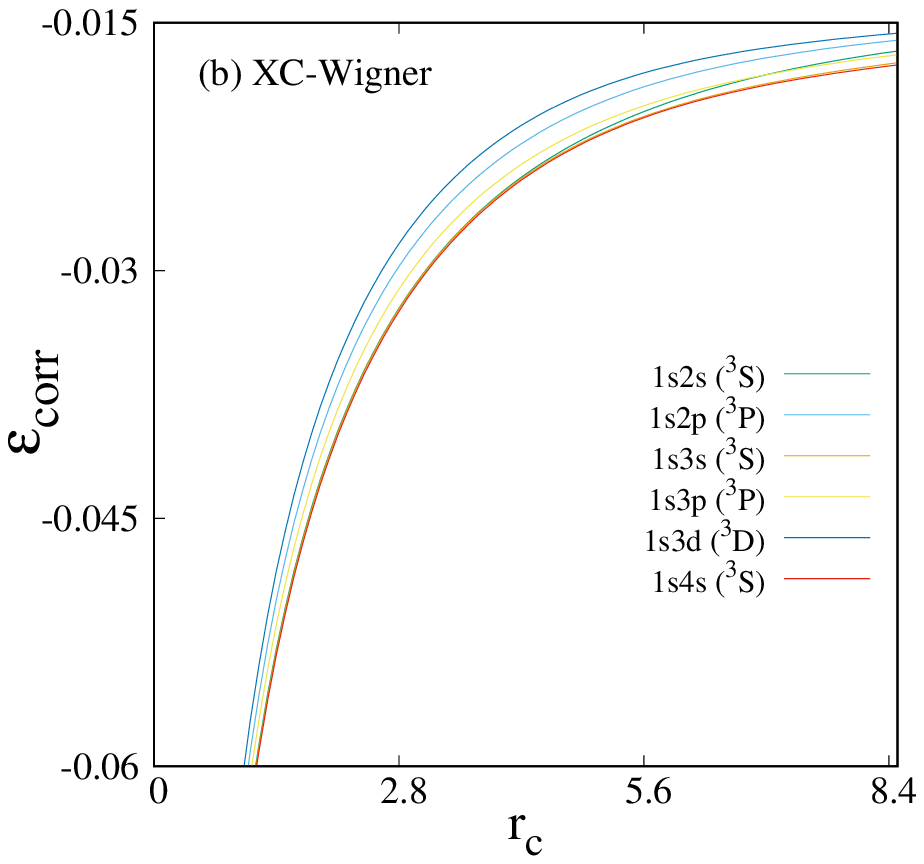}
\end{minipage}%
\caption{\label{fig:figure2}Correlation energies, with changes in $r_c$, for singly excited states of confined H$^{-}$, in panels 
(a) singlet (b) triplet states for Wigner correlation, while (c), (d) give same for LYP functional.}
\end{figure}
             
\begin{figure}                         
\begin{minipage}[c]{0.50\textwidth}\centering
\includegraphics[scale=0.8]{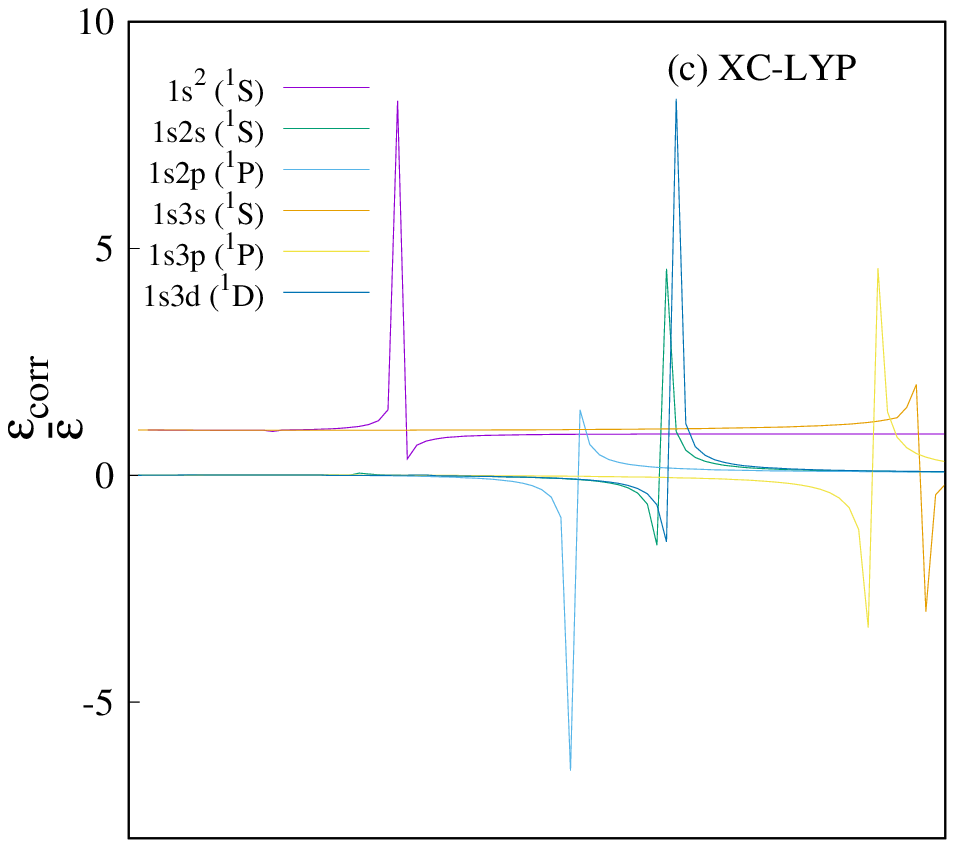}
\end{minipage}%
\begin{minipage}[c]{0.50\textwidth}\centering
\includegraphics[scale=0.8]{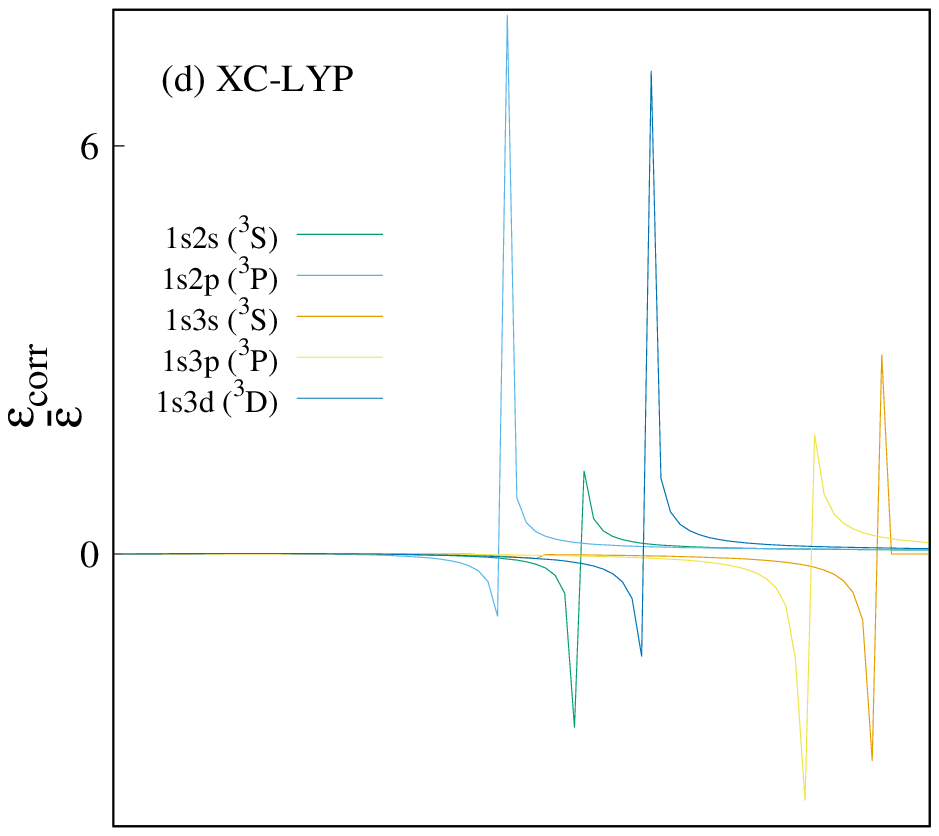}
\end{minipage}%
\vspace{5mm}
\hspace{0.2in}
\begin{minipage}[c]{0.50\textwidth}\centering
\includegraphics[scale=0.85]{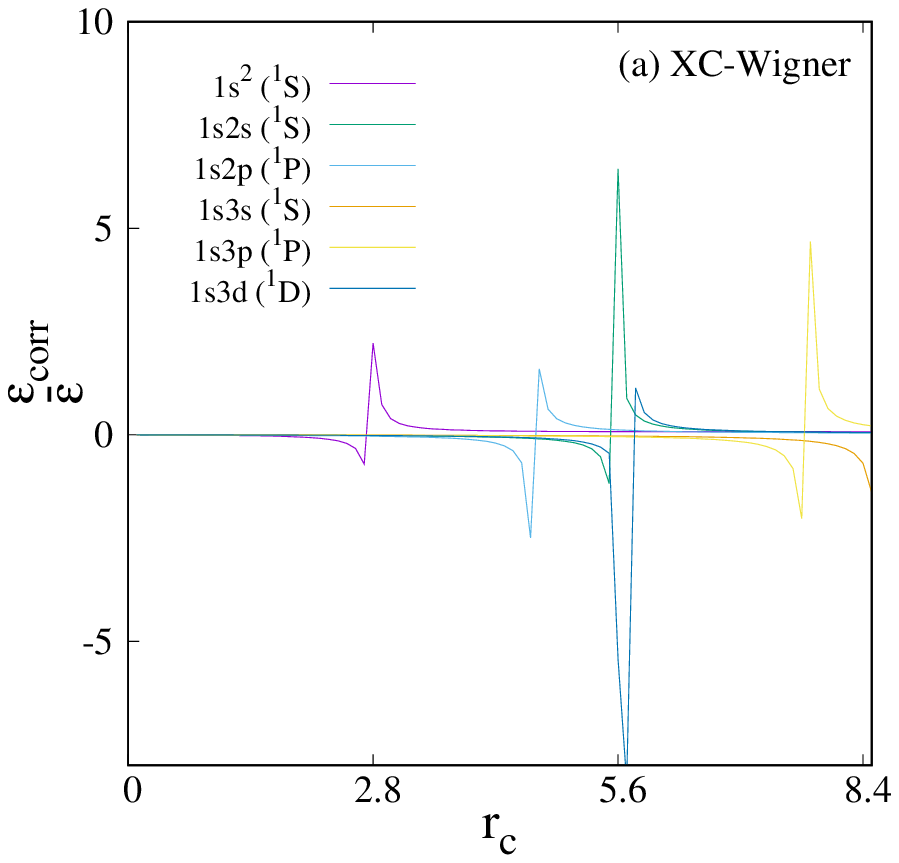}
\end{minipage}%
\begin{minipage}[c]{0.50\textwidth}\centering
\includegraphics[scale=0.85]{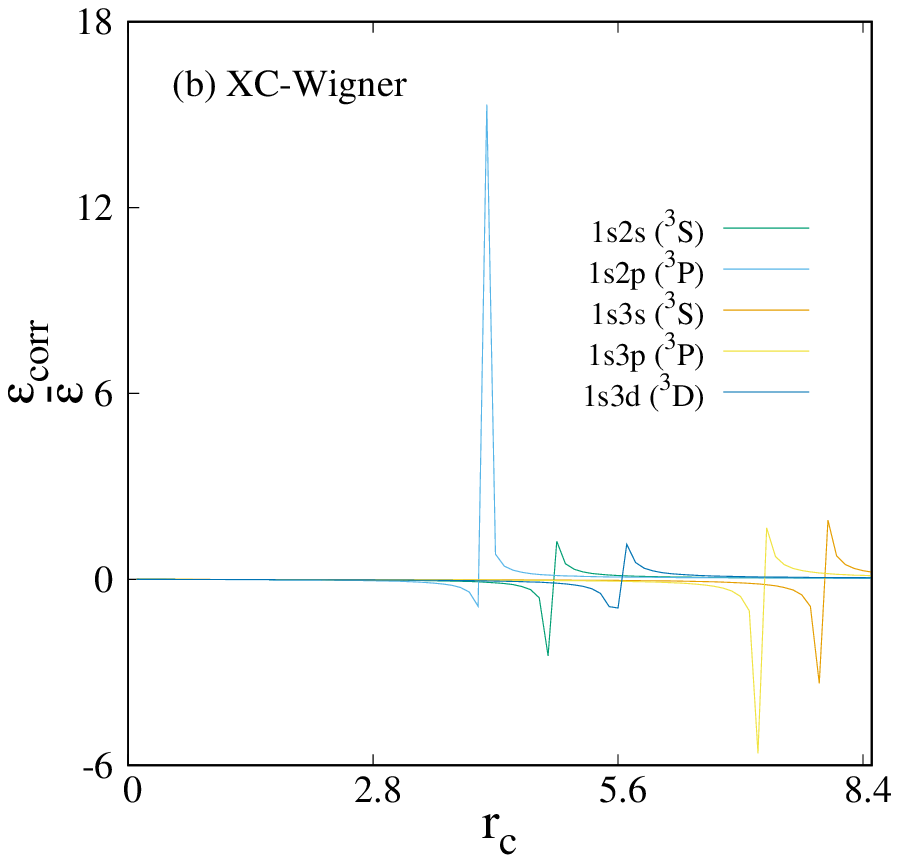}
\end{minipage}%
\caption{\label{fig:figure3} $\left(\frac{\epsilon_{corr}}{\epsilon}\right)$ for singly excited states of confined H$^{-}$, w.r.t.
box size, in panels (a) singlet (b) triplet states for Wigner correlation, while (c), (d) give these for LYP functional.}
\end{figure}

These results of Tables~\ref{tab:table1}--\ref{tab:table3} encourage us to investigate the impact of 
confinement on excited states in a qualitative manner. Therefore, such energies are 
plotted in panels (a) and (b) of Fig.~\ref{fig:figure1} for a few singlet and triplet singly excited states as a function of
$r_{c}$. In addition to the states investigated in above three tables, in this occasion, we have considered some additional states, 
such as, 1s3s and 1s4s $^{3,1}$S, 1s3p $^{3,1}$P and 1s3d $^{3,1}$D. In order to get a better insight about the 
crossing amongst various states, an amplified region ($\epsilon \le 0.2$) of panel (a) is demonstrated in panel (b)
with improved resolution. X-only, XC-Wigner, XC-LYP energies generate qualitatively resembling plots. Hence, 
we take liberty to use X-only energies to point out some general features. In a free H$^{-}$ ion the possible ordering 
of states under consideration is: $\epsilon_{\mathrm{1s4s}}(^1\mathrm{S})>$ 
$\epsilon_{\mathrm{1s4s}}(^3\mathrm{S})>$ $\epsilon_{\mathrm{1s3d}}(^1\mathrm{D})>$ $\epsilon_{\mathrm{1s3d}}(^3\mathrm{D})>$
$\epsilon_{\mathrm{1s3p}}(^1\mathrm{P})>$ $\epsilon_{\mathrm{1s3p}}(^3\mathrm{P})>$ $\epsilon_{\mathrm{1s3s}}(^1\mathrm{S})>$ 
$\epsilon_{\mathrm{1s3s}}(^3\mathrm{S})>$ $\epsilon_{\mathrm{1s2p}}(^1\mathrm{P})>$ $\epsilon_{\mathrm{1s2p}}(^3\mathrm{P})>$ 
$\epsilon_{\mathrm{1s2s}}(^1\mathrm{S})>$ $\epsilon_{\mathrm{1s2s}}(^3\mathrm{S})>$ $\epsilon_{\mathrm{1s}^2}(^1\mathrm{S})$.
It has been found earlier that the influence of confinement seems to be more pronounced on valence orbitals leading to the rearrangement 
of atomic states at strong confinement regime \textcolor{red}{\cite{montgomery15, majumdar20a}}. Once we move from free to confinement 
limit, multiple crossover between states occur, and the above order gets dissolved. This ordering is a function of $r_{c}$. From panel (a) it can
be checked that at $r_{c}=4,~5.6,~7.2$ and 8.1 crossover between 1s2s $^{1}$S, 1s3d $^{1}$D; 1s2s $^{1}$S, 1s3d $^{3}$D;
1s2s $^{3}$S, 1s2p $^{3}$P; 1s2s $^{3}$S, 1s2p $^{1}$P occur respectively. Moreover, the last three crossings are clearly visible from 
panel (b). It is to be noted that, beyond the range of $r_c$ plotted here, several other crossings happen, 
which are not given this figure, to avoid clumsiness.

The outcomes of Fig.~\ref{fig:figure1} motivate us to explore the ordering of various singly excited singlet and triplet states 
at strong confinement region. In this context, the energies for first thirty two singly excited triplet and singlet states
arising from different singly excited configurations are provided in ascending order at $r_{c}=0.1$ in Table~\ref{tab:table4}. 
The third, fourth and fifth columns provide the X-only, XC-Wigner and XC-LYP results respectively. It has been verified thoroughly that, 
apart from the presented states, no intermediate singly excited state can be found to lie in between them. \textcolor{red}{It is to be 
noted here that, in this limit of confinement, for all the states under consideration, Hund's rule is satisfied; singlet states 
possess higher energy than the triplets.}

\begingroup           
\squeezetable
\begin{table}
\caption {\label{tab:table5} $S_{\rvec}, S_{\pvec}$ in ground state of confined H$^-$. See text for details.}
\centering
\begin{ruledtabular} 
\begin{tabular}{c | c c c c  c c  c c}
 $r_c$ & \multicolumn{2}{c}{X-only} & \multicolumn{2}{c}{XC-Wigner} & \multicolumn{2}{c}{XC-LYP} & 
\multicolumn{2}{c}{Literature}             \\
\cline{2-3} \cline{4-5} \cline{6-7} \cline{8-9}  
& $S_{\rvec}$ & $S_{\pvec}$  & $S_{\rvec}$ & $S_{\pvec}$ & $S_{\rvec}$ & $S_{\pvec}$  & $S_{\rvec}$$^{(a)}$ & $S_{\pvec}$$^{(a)}$  \\
\hline
 0.1    &  $-$6.2407   & 12.847   &  $-$6.2407  &  12.8500    &  $-$6.2407 &  12.8500   &  -           &  -            \\
 0.2    &  $-$4.1701   & 10.7750  &  $-$4.1702  &  10.7750    &  $-$4.1701 &  10.7750   &  -           &  -           \\
 0.3    &  $-$2.9628   & 9.5621   &  $-$2.9629  &  9.5621     &  $-$2.9628 &  9.5621    &  $-$2.948    &  9.394      \\
 0.5    &  $-$1.4493   & 8.0373   &  $-$1.4495  &  8.0373     &  $-$1.4493 &  8.0373    &  $-$1.423    &  7.954      \\
 0.8    &  $-$0.06968  & 6.6413   &  $-$0.07027 &  6.6416     & $-$0.06977 &  6.6413    &  $-$0.032    &  6.593      \\
 1      &  0.5781      & 5.9832   &  0.5771     &  5.9838     &  0.5779    &  5.9833    &  0.6209      &  5.943      \\
 1.2    &  1.1023      & 5.4494   &  1.1007     &  5.4503     &  1.1020    &  5.4495    &  1.1486      &  5.414      \\
 1.5    &  1.7353      & 4.8034   &  1.7327     &  4.8051     &  1.7347    &  4.8039    &  -           &  -          \\
 1.8    &  2.2430      & 4.2848   &  2.2391     &  4.2874     &  2.2421    &  4.2854    &  2.2912      &  4.255      \\
 2      &  2.5313      & 3.9904   &  2.5263     &  3.9940     &  2.5300    &  3.9913    &  2.5777      &  3.963      \\
 2.5    &  3.1253      & 3.3872   &  3.1171     &  3.3939     &  3.1230    &  3.3890    &  3.1633      &  3.366      \\
 3      &  3.5883      & 2.9228   &  3.5760     &  2.9337     &  3.5843    &  2.9263    &  3.6133      &  2.913      \\
 3.5    &  3.9583      & 2.5570   &  3.9409     &  2.5731     &  3.9517    &  2.5629    &  3.9672      &  2.563      \\
 4      &  4.2583      & 2.2653   &  4.2350     &  2.2876     &  4.2479    &  2.2750    &  4.2494      &  2.29       \\
 5      &  4.7051      & 1.8433   &  4.6675     &  1.8801     &  4.6818    &  1.8652    &  4.6607      &  1.908      \\
 7      &  5.2132      & 1.3942   &  5.1425     &  1.4627     &  5.1432    &  1.4570    &  5.1139      &  1.522      \\
 8      &  5.3524      & 1.2811   &  5.2657     &  1.3636     &  5.2568    &  1.3649    &  5.2423      &  1.423      \\
10      &  5.5097      & 1.1623   &  5.3958     &  1.2666     &  5.3769    &  1.2755    &  -           &  -           \\
15      &  5.6180      & 1.0911   &  5.4709     &  1.2180     &  5.4494    &  1.2309    &  -           &  -           \\
25      &  5.6300      & 1.0852   &  5.4753     &  1.2162     &  5.4507    &  1.2330    &  -           &  -           \\
\end{tabular}
\end{ruledtabular}
\begin{tabbing}
$^{\mathrm{(a)}}$Ref.~\cite{sen05}. \hspace{25pt}  \=
\end{tabbing}
\end{table}
\endgroup       

\begingroup           
\squeezetable
\begin{table}
\caption {\label{tab:table6} $S_{\rvec}, S_{\pvec}$ in 1s2s $^3$S and 1s2p $^3$P states of confined H$^-$. See text for details.}
\centering
\begin{ruledtabular} 
\begin{tabular}{c | c c | c c | c c | c c | c c | c c }
       & \multicolumn{6}{c|}{1s2s $^{3}$S} & \multicolumn{6}{c}{1s2p $^{3}$P} \\
\cline{2-8} \cline{8-13}
 $r_c$ & \multicolumn{2}{c}{X-only} & \multicolumn{2}{c}{XC-Wigner} & \multicolumn{2}{c|}{XC-LYP}   
       & \multicolumn{2}{c}{X-only} & \multicolumn{2}{c}{XC-Wigner} & \multicolumn{2}{c}{XC-LYP}  \\
\cline{2-3} \cline{4-5} \cline{6-7} 
\cline{8-9} \cline{10-11} \cline{12-13} 
   & $S_{\rvec}$ & $S_{\pvec}$  & $S_{\rvec}$ & $S_{\pvec}$ & $S_{\rvec}$ & $S_{\pvec}$  
   & $S_{\rvec}$ & $S_{\pvec}$  & $S_{\rvec}$ & $S_{\pvec}$ & $S_{\rvec}$ & $S_{\pvec}$ \\
\hline
 0.1    &  $-$6.21007   & 14.1888  &  $-$6.21007 &  14.1888  &  $-$6.21007  &  14.1888 
        &  $-$6.16304   & 13.0692  &  $-$6.16304 &  13.0711  &  $-$6.16304  &  13.0708 \\
 0.2    &  $-$4.13691   & 12.1115  &  $-$4.13691 &  12.1113  &  $-$4.1369   &  12.1112   
        &  $-$4.08845   & 10.9927  &  $-$4.08846 &  10.9927  &  $-$4.0884   &  10.9927  \\   
 0.3    &  $-$2.92690   & 10.8967  &  $-$2.92692 &  10.8964  &  $-$2.9269   &  10.8977   
        &  $-$2.87704   &  9.7777  &  $-$2.87707 &   9.7722  &  $-$2.8770   &  9.7774        \\
 0.5    &  $-$1.40751   &  9.3668  &  $-$1.40758 &   9.3655  &  $-$1.4075 &  9.3671        
        &  $-$1.35498   &  8.2477  &  $-$1.35507 &   8.2474  &  $-$1.3549   &  8.2474         \\ 
 0.7    &  $-$0.41162   &  8.3593  &  $-$0.41177 &   8.3593  &  $-$ 0.4116 & 8.3592         
        &  $-$0.35658   &  7.2412  &  $-$0.35680 &   7.2411  &  $-$0.3566  & 7.2412           \\
 1      &     0.63727   &  7.2955  &   0.63690   &   7.2956  &  0.63721   &  7.2948        
        &     0.69568   &  6.1776  &   0.6951    &   6.1779  &  0.69559   &  6.1777          \\
 1.2    &     1.16957   &  6.7534  &   1.16902   &   6.7535  &  1.1694    &  6.7527         
        &     1.22992   &  5.6364  &   1.229152  &   5.6368  &  1.2297    &  5.6365          \\
 1.5    &     1.81613   &  6.0918  &   1.81523   &   6.0923  &  1.81598   &  6.0923        
        &     1.87883   &  4.9779  &   1.87755   &   4.9785  &  1.8785    &  4.9781          \\
 1.8    &     2.33915   &  5.5559  &   2.33780   &   5.5565  &  2.33891   &  5.5561        
        &     2.40340   &  4.4451  &   2.40149   &   4.4462  &  2.4029    &  4.4453          \\
 2      &     2.63865   &  0.1578  &   2.63697   &   5.2493  &  2.6383    &  5.2485        
        &     2.70343   &  4.1406  &   2.70102   &   4.1421  &  2.7028    &  4.1409          \\
 2.5    &     3.26448   &  4.6048  &   3.26181   &   4.6065  &  3.2639    &  4.6051        
        &     3.32850   &  3.5086  &   3.32455   &   3.5114  &  3.3275    &  3.5092          \\ 
 3      &     3.76475   &  4.0903  &   3.76088   &   4.0929  &  3.76398   &  4.0907        
        &     3.82454   &  3.0127  &   3.43221   &   3.4032  &  3.8228    &  3.0139     \\
 5      &     5.08159   &  2.7464  &   5.07112   &   2.7555  &  5.0747    &  2.7515        
        &     5.08496   &  1.8241  &   5.06728   &   1.8428  &  5.0721    &  1.8357   \\
 6      &     5.50678   &  2.3211  &   5.492351  &   2.3341  &  5.4877    &  2.3350        
        &     5.46497   &  1.5157  &   5.44087   &   1.5426  &  5.4369    &  1.5381   \\
 8      &     6.11312   &  1.7284  &   6.090808  &   1.7495  &  6.0292    &  1.7884        
        &     5.9898    &  1.1630  &   5.95727   &   1.2011  &  5.9018    &  1.2115   \\
10      &     6.52773   &  1.3347    & 6.4987    &  1.3619   &  6.3378    &  1.4698      
        &     6.3630    &  0.9543    & 6.3269    &  0.9964   &  -         &  -  \\ 
15      &     7.18923   &  0.7177    & 7.1486    & 0.7528    &  -         &  -      
        &     7.0703    &  0.5437    & 6.9781    &  0.6208   &  -         &  -       \\ 
25      &     7.9460    & $-$0.05128 & 7.8851    & 0.0932    &  -         &  -             
        &     7.8138    & $-$0.00327 & 7.7680    & 0.04055   &  -         &  -                 \\ 
\end{tabular}
\end{ruledtabular}
\end{table}
\endgroup                          

Till now we were involved in exploring the impact of confinement on total energy of H$^{-}$ ion. It has been found that, 
both X-only and correlated energies diminish with increase in $r_{c}$ and merge to respective free limits. At this point, 
it is sensible to investigate the 
variation in correlation with change in confinement strength. In this regard, Wigner and LYP correlation energies are 
plotted in panels (a), (b) and (c), (d) of Fig.~\ref{fig:figure2} respectively. The left panels (a), (c) represent 
corresponding singlet states and right panels (b), (d) indicate respective triplet states. From (a) and (b) 
it is evident that for both singlet and triplet states Wigner correlation energies accelerate with growth in $r_{c}$.
This observation corroborates the pattern from a Hylleraas calculation \cite{wilson10}. Moreover, through out
the range in $r_{c}$ the Wigner correlation energies remain \emph{negative} for all these states. It is noteworthy 
that, correlation energy is the difference between \emph{exact} energy and HF energy. Further, HF energy is always upper
bound to the \emph{exact} energy. Therefore, correlation energy should be \emph{negative}. Here, Wigner correlation energies obey
this criteria. Now we move to analyze the LYP correlation energy. On the contrary, from panels (c) and (d) it can be seen that,
except for 1s4s $^{3,1}$S states, LYP correlation energies for all other states decrease with progress in $r_{c}$. However, 
for 1s4s $^{3,1}$S states it initially decreases with $r_{c}$, attains a flat minimum and then increases again. However, in the entire 
range of $r_{c}$ such energies for all these given states attain both positive as well as negative values.

We have also examined the ratio of correlation energy and respective total energy in Fig.~\ref{fig:figure3} 
for Wigner (panels (a), (b)) and LYP (panels (c),(d)) functionals for the same set of excited states studied in 
Fig.~\ref{fig:figure2}. Here also panels (a), (c) refer to singlet states and (b), (d) signify triplet
states. For each of these states involving either of the functionals, a sudden jump occurs at a characteristic $r_{c}$. This jump 
is not due to the sign change in either of the energies. Because, Wigner correlation energies are always negative, but LYP can be both 
positive and negative. Thus, for all these 
states these two energies connected to XC-Wigner changes their domain from negative correlation energy and positive total 
energy in low $r_{c}$ 
to negative correlation energy and negative total energy in free limit. However, for XC-LYP case, same change occurs from positive 
correlation energy and positive total energy region to negative correlation energy and negative total energy. 

\subsection{Shannon entropy and Onicescu energy}
Now we apply this method to compute $S, E$ in conjugate $r$ and $p$ spaces. This gives a scope to verify
and assess the quality of density in such states under hard confinement. Because, they act as descriptor in
interpreting various chemical phenomena. Moreover, it will help us to understand the correlation contribution in 
present endeavor.  

\begin{figure}                         
\begin{minipage}[c]{0.33\textwidth}\centering
\includegraphics[scale=0.6]{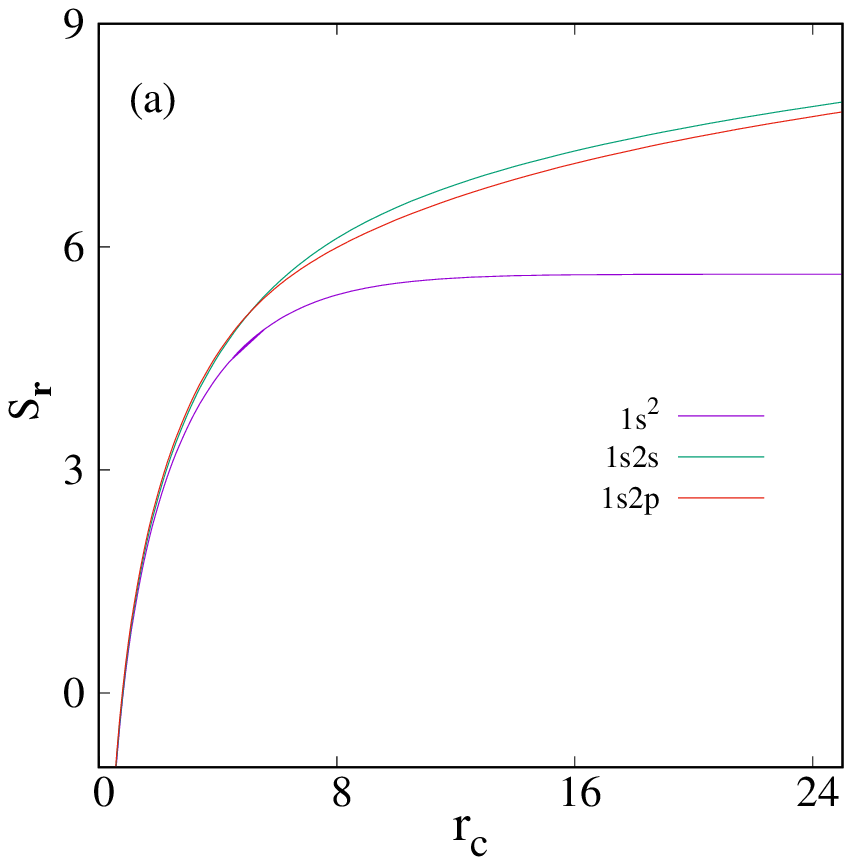}
\end{minipage}%
\begin{minipage}[c]{0.33\textwidth}\centering
\includegraphics[scale=0.6]{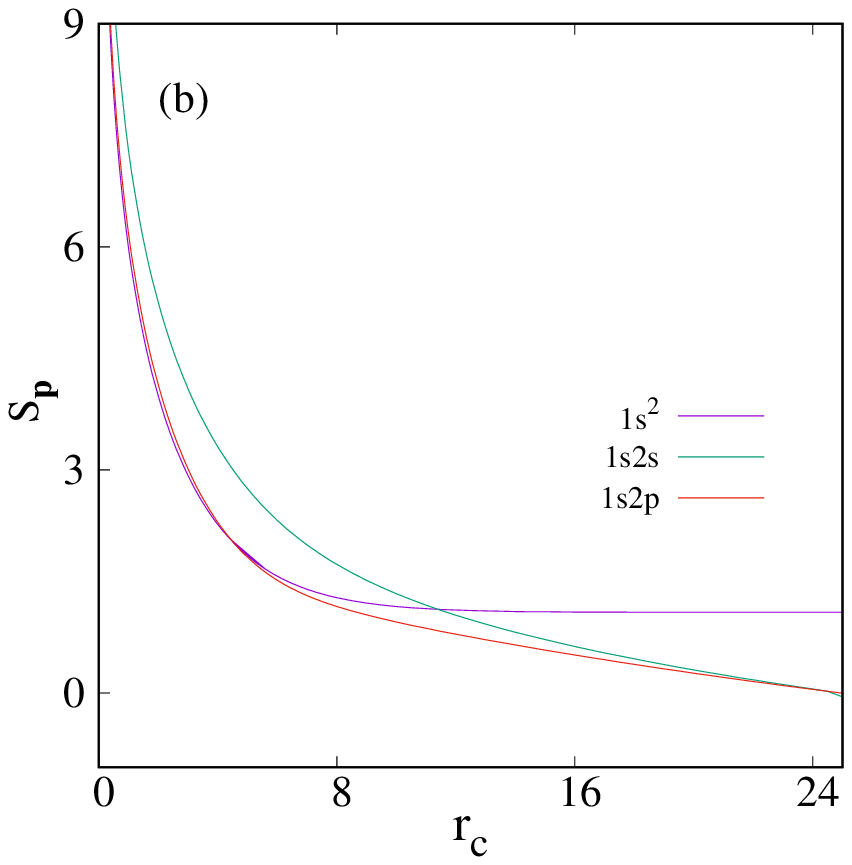}
\end{minipage}%
\begin{minipage}[c]{0.33\textwidth}\centering
\includegraphics[scale=0.6]{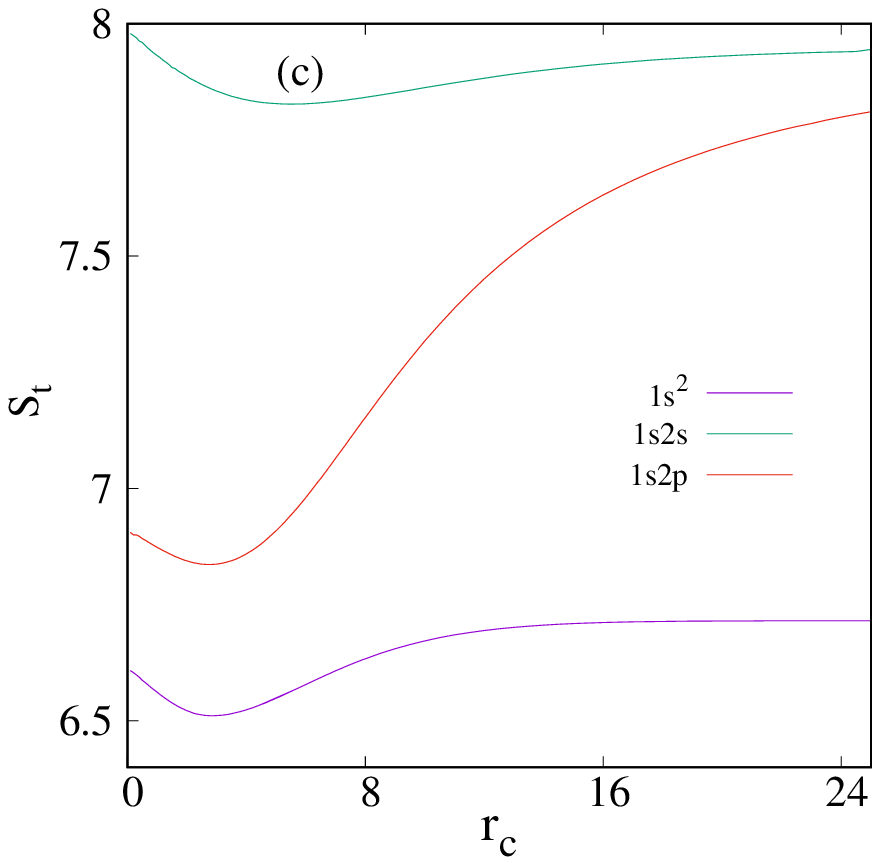}
\end{minipage}%
\caption{\label{fig:figure4} Variation of (a)$S_{\rvec}, (b)S_{\pvec}, (c)S_{t}$ with change in $r_{c}$ for H$^{-}$ ion.
See text for details.}
\end{figure} 

Table~\ref{tab:table5} tabulates the numerical values of $S_{\rvec}$ and $S_{\pvec}$ for H$^{-}$ ion in ground state at certain
selected $r_{c}$ values. In all three occasions (X-only, XC-Wigner, XC-LYP) $S_{\rvec}$ increases and $S_{\pvec}$ decreases with 
growth in $r_{c}$. This
amplifies the conclusion that, electron density gets compressed with strengthening of confinement effect. At strong confinement 
zone, both X-only and correlated results in either space become identical. However, with increase in $r_{c}$ this situation 
alters indicating the contribution of correlation effect in density. Similar observation was also reported in \cite{majumdar20} 
for He-isoelectronic series involving He, Li$^{+}$ and Be$^{+}$. At $r_{c} \ge 1$ regime, X-only values of $S_{\rvec}$ are higher 
compared to both Wigner and LYP results. However, in $p$-space, a reverse behavior is noticed. X-only values are smaller relative 
to correlated results. The BLYP \cite{sen05} $S_{\rvec}$ and $S_{\pvec}$ are quoted in last two columns of 
table. The Wigner and LYP Shannon entropies are in complete correspondence with these cited values. 
It is evident that, $(S_{\rvec}+S_{\pvec})$ always remains greater that its limiting value of $3(1+\ln \pi)$.   

\begin{figure}                         
\begin{minipage}[c]{0.5\textwidth}\centering
\includegraphics[scale=0.90]{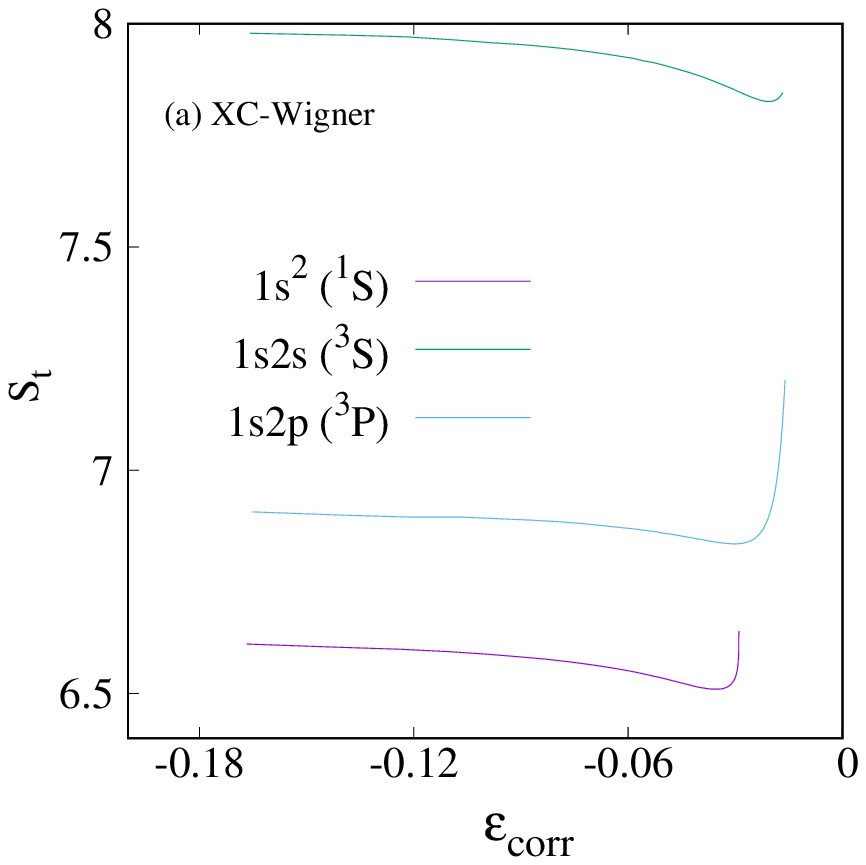}
\end{minipage}%
\begin{minipage}[c]{0.5\textwidth}\centering
\includegraphics[scale=0.90]{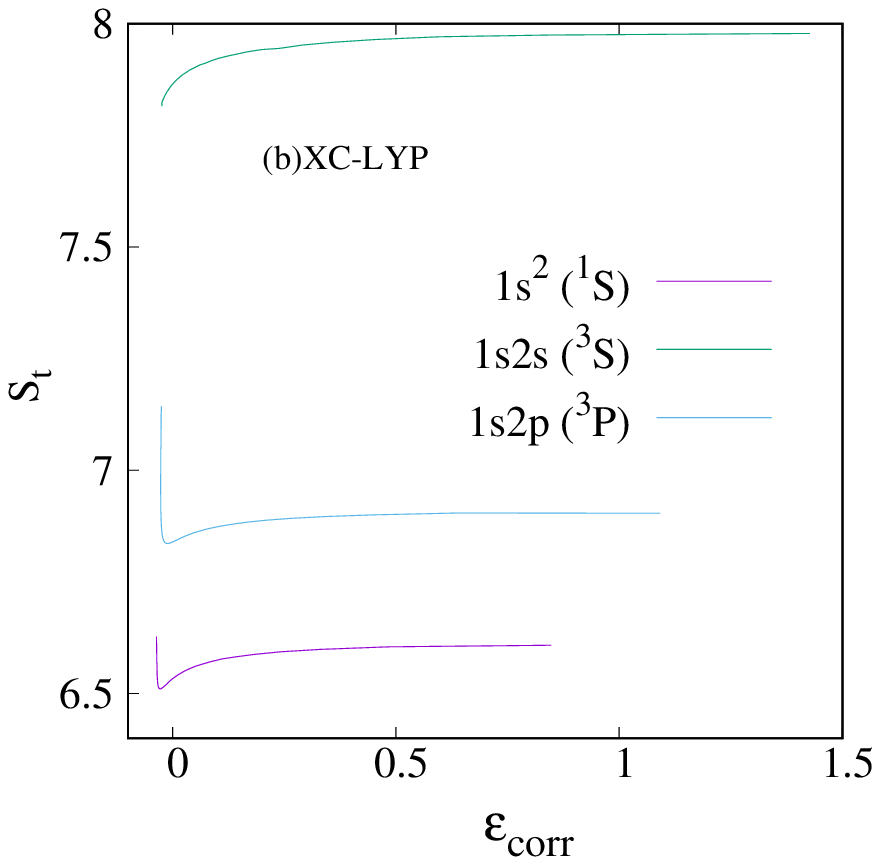}
\end{minipage}%
\caption{\label{fig:figure5} Variation of $S_{t}$ with change in correlation energy employing (a) XC-Wigner and (b) XC-LYP functional.
See text for details.}
\end{figure} 

Next, $S_{\rvec}$, $S_{\pvec}$ values for 1s2s $^{3}$S and 1s2p $^{3}$P states are reported in Table~\ref{tab:table6}. The 
second to seventh columns represent $^{3}$S, whereas last six columns indicates $^{3}$P. No previous 
literature is available to match with our results. Analogous to the ground state, here also for either of the states, X-only and 
correlated entropies are uniform at strong confinement zone and the correlation contribution grows with 
rise in $r_{c}$. Further, at $r_{c} \ge 1$ region, the X-only results are comparatively higher than those from Wigner and LYP. 
It has been found that, at low to moderate $r_{c}$ region energies of 1s2s $^{3}$S state are larger than 
1s2p $^{3}$P, and crossing occurs when $r_{c}$ value lies in between $5$ and $6$. However, an exactly opposite 
trend is encountered here in the 
context of $S_{\rvec}$. These values for the former state is less than that of the latter, and a crossover happens at 
$r_{c} \approx 6$. However, $S_{\pvec}$ obeys the same pattern as observed in energy. As usual the sum of $r$- and $p$-space
$S$ is higher than the bound value of 6.43418.                  

\begingroup           
\squeezetable
\begin{table}
\caption {\label{tab:table7} $E_{\rvec}, E_{\pvec}$ in ground state of confined H$^-$. See text for details.}
\centering
\begin{ruledtabular} 
\begin{tabular}{c | c c c c  c c }
 $r_c$ & \multicolumn{2}{c}{X-only} & \multicolumn{2}{c}{XC-Wigner} & \multicolumn{2}{c}{XC-LYP} \\
\cline{2-3} \cline{4-5} \cline{6-7} 
  & $E_{\rvec}$ & $E_{\pvec}$  & $E_{\rvec}$ & $E_{\pvec}$ & $E_{\rvec}$ & $E_{\pvec}$  \\
\hline
 0.1    &  681.5854   & 0.000003 &  681.5874  &  0.000003 & 681.5855  & 0.000003   \\
 0.2    &  86.4424    & 0.00003  &  86.4442   &  0.00003 & 86.4426  & 0.00003   \\
 0.3    &  25.9984    & 0.0001   &  26.00007  &  0.0001 & 25.9985    & 0.0001   \\
 0.5    &  5.7942    & 0.0004    &  5.7956   &  0.0004 & 5.7944   & 0.0004   \\
 0.8    &  1.4880    & 0.0019    &  1.4892   &  0.0019  & 1.4882   & 0.0019    \\
 1      &  0.7901    & 0.0037  &  0.7911   &  0.0037  & 0.7902   & 0.0037    \\
 1.2    &  0.4751    & 0.0064  &  0.4761   &  0.0064  &  0.4753  & 0.0064    \\
 1.5    &  0.2588    & 0.0122   &  0.2597   &  0.0122   &  0.2590  & 0.0122     \\
 1.8    &  0.1601    & 0.0206   &  0.1610   &  0.0206   &  0.1396  & 0.0240     \\
 2      &  0.1224    & 0.0278   &  0.1232   &  0.0277   &  0.1226  & 0.0278     \\
 2.5    &  0.0713    & 0.0518   &  0.0721   &  0.0515   &  0.0715  & 0.0517     \\
 3      &  0.0476    & 0.0847   &  0.0484   &  0.0840   &  0.0479  & 0.0845     \\
 3.5    &  0.0350    & 0.1266   &  0.0358   &  0.1250   &  0.0353  & 0.1260     \\
 4      &  0.0277    & 0.1766   &  0.0285   &  0.1736   &  0.0281  & 0.1753     \\
 5      &  0.0202    & 0.2954    &  0.0211   &  0.2872   &  0.0207  &  0.2903     \\
 7      &  0.0151    & 0.5624    &  0.0162   &  0.5296   &  0.0160  &  0.5277     \\
 8      &  0.0142    & 0.6862    &  0.0154   &  0.6350    &  0.0153  &  0.6257    \\
10      &  0.0135    & 0.8837    &  0.0147   &  0.7898    &  0.0147  &  0.7681     \\
15      &  0.0132    & 1.1022    &  0.0145   &  0.9284    &  0.0144  &  0.9017      \\
25      &  0.0132    & 1.1445    &  0.0143   &  0.9443    &  0.0145  &  0.9179              \\
\end{tabular}
\end{ruledtabular}
\end{table}
\endgroup          
    
In order to get a better insight about the influence of confinement on entropies, $S_{\rvec}$, $S_{\pvec}$, $S_{t}$ are plotted 
as function of $r_{c}$, in panels (a)--(c) of Fig.~\ref{fig:figure4}, for 1s$^{2}$ $^{1}$S, 1s2s $^{3}$S, 1s2p $^{3}$P states. 
The correlation effect does not alter the qualitative nature of the graph. Hence, X-only results suffice. As seen, 
$S_{\rvec}$ progresses with gain in $r_{c}$, while $S_{\pvec}$ declines. In conformity with Table~VI, panel (a) 
also indicates the crossover between 1s2s $^{3}$S, 1s2p $^{3}$P states at around $r_{c}=6$. 
However, multiple crossovers between 1s$^{2}$ $^{1}$S, 1s2s $^{3}$S; 1s$^{2}$ $^{1}$S, 1s2p $^{3}$P are seen at 
$r_{c}\approx 6, 11$ respectively in panel (b). $S_{t}$ in all these three cases, initially decline, then attain 
a minimum and finally increase.    

In one-electron system, $S_{t}$ is independent of effective nuclear charge $Z$. However, in a many-electron
system, this situation alters and it depends on $Z$. Previously $S_{t}$ was employed in explaining the correlation effect
in both free and confined conditions \cite{guevara03}. Now, $S_{t}$ has been plotted as a function of correlation 
energy in Fig.~\ref{fig:figure5}. Panels (a)-(b) represent Wigner and LYP functionals. In the former case, for 
all these three states, it decays with rise in $\epsilon_{corr}$, then reaches a minimum and then sharply increases thereafter. On
the contrary, for LYP functional involving 1s$^{2}$ $^{1}$S, 1s2s $^{3}$S states, it sharply decreases to a minimum and 
then gradually increases with rise in $\epsilon_{corr}$. Further, for 1s2p $^{3}$P state, it always rises with $\epsilon_{corr}$.  

\begingroup           
\squeezetable
\begin{table}
\caption {\label{tab:table8} $E_{\rvec}, E_{\pvec}$ in 1s2s $^3$S and 1s2p $^3$P states of confined H$^-$. See text for details.}
\centering
\begin{ruledtabular} 
\begin{tabular}{c | c c| c c | c c | c c | c c | c c }
 & \multicolumn{6}{c|}{1s2s $^{3}$S} & \multicolumn{6}{c}{1s2p $^{3}$P} \\
\cline{2-8} \cline{8-13}
 $r_c$ & \multicolumn{2}{c}{X-only} & \multicolumn{2}{c}{XC-Wigner} & \multicolumn{2}{c|}{XC-LYP}   
       & \multicolumn{2}{c}{X-only} & \multicolumn{2}{c}{XC-Wigner} & \multicolumn{2}{c}{XC-LYP}  \\
\cline{2-3} \cline{4-5} \cline{6-7} 
\cline{8-9} \cline{10-11} \cline{12-13} 
   & $E_{\rvec}$ & $E_{\pvec}$  & $E_{\rvec}$ & $E_{\pvec}$ & $E_{\rvec}$ & $E_{\pvec}$  
   & $E_{\rvec}$ & $E_{\pvec}$  & $E_{\rvec}$ & $E_{\pvec}$ & $E_{\rvec}$ & $E_{\pvec}$ \\
\hline
 0.1    &  898.6345    & 1$\times 10^{-8}$   &  898.6358 &   1$\times 10^{-8}$  &  898.6346  &      1$\times 10^{-8}$ 
        &  931.7284    & 3$\times  10^{-7}$  &  931.7299 &   3 $\times 10^{-7}$ &  931.7285  &      3$\times 10^{-7}$     \\
 0.2    &  114.0409    & 1$\times 10^{-7}$   &  114.0421 &   1$\times 10^{-7}$  & 114.0410   &     1$\times 10^{-7}$   
        &  117.3172    & 2$\times 10^{-7}$   &  117.3185 &   2$\times 10^{-7}$  & 117.3173   &     2$\times 10^{-7}$   \\
0.3    &   34.3147    &  3$\times 10^{-7}$   &  34.3158  &  3$\times 10^{-7}$  &  34.3148   &    3$\times 10^{-7}$         
       &   35.0239    &  9$\times 10^{-7}$   &  35.0251  & 9$\times 10^{-7}$   &  35.0240   &    9$\times 10^{-7}$  \\
0.5    &    7.6508    &  1$\times 10^{-6}$   &   7.6517  &  1$\times 10^{-6}$   &  7.6509    &   1$\times 10^{-6}$        
       &    7.6867    &  4$\times 10^{-6}$   &   7.6876  &  4$\times 10^{-6}$   &  7.6868    &   4$\times 10^{-6}$  \\
0.7    &    2.8815    &  4$\times 10^{-6}$   &   2.8823  &  4$\times 10^{-6}$    &  2.8816    &  4$\times 10^{-6}$        
       &    2.8497    &  12$\times 10^{-5}$  &   2.8505  &  12$\times 10^{-5}$    &  2.8498    &  12$\times 10^{-5}$  \\
1      &    1.0408    &  13$\times 10^{-5}$  &   1.0415  &  13$\times 10^{-5}$    &  1.0409    &  13$\times 10^{-5}$         
       &    1.0055    &  3$\times 10^{-5}$   &   1.0062 &  3$\times 10^{-5}$    &  1.0056   &  3$\times 10^{-5}$            \\
1.2    &    0.6244    &  2$\times 10^{-5}$   &   0.6250 &  2$\times 10^{-5}$    &  0.6245    &  2$\times 10^{-5}$         
       &    0.5941    &  6$\times 10^{-5}$   &   0.5947 &  6$\times 10^{-5}$    &  0.5942    &  6$\times 10^{-5}$            \\
1.5    &    0.3382    &  4$\times 10^{-5}$   &   0.3388 &  4$\times 10^{-5}$    &  0.3384 &  4$\times 10^{-5}$        
       &    0.3148    &  0.0116              &   0.3153 & 0.0116                & 0.3149  &  0.0116           \\
1.8    &    0.2077    & 8$\times 10^{-5}$    &   0.2082 & 8$\times 10^{-5}$     & 0.2078 &  8$\times 10^{-5}$        
       &    0.1893    & 0.0198               &   0.1898 & 0.0197    &  0.1894    & 0.0198            \\
2      &    0.1578    & 1$\times 10^{-4}$    &   0.1583 & 1$\times 10^{-4}$    &    0.1579 &  1$\times 10^{-4}$        
       &    0.1420    & 0.0268               &   0.1425 & 0.0267               &  0.1421  &   0.0268          \\  
2.5    &    0.0900    & 0.0212               &   0.0905 & 0.0212               &  0.0901  &   0.0212        
       &    0.0789    & 0.0504               &   0.0794 & 0.0502               &  0.0790  &   0.0503          \\
 3      &   0.0585    & 0.0365               &   0.0590 & 0.0364               &  0.0586  &   0.0365        
        &   0.0504    & 0.0832               &   0.0719 & 0.0560               &  0.0506  &   0.0830           \\
 5      &   0.0215    & 0.1635               &   0.0219 & 0.1630               &  0.0217  &   0.1631        
        &    0.0199   & 0.2972      &   0.0204  & 0.2924      & 0.0202     & 0.2939     \\
 6      &    0.0166    & 0.2771     &   0.0170 & 0.2760     & 0.0169    & 0.2741         
        &    0.0171    & 0.4352     &   0.0177 & 0.4254     & 0.0175    & 0.4248        \\
 8      &    0.0126    & 0.6312     &   0.0130 & 0.6273     & 0.0132    & 0.5892         
        &    0.0162    & 0.7373     &   0.0170 & 0.7149     & 0.0166    & 0.6862     \\
10      &    0.0112   & 1.1891   & 0.0169  &  1.1789       & 0.0121    & 0.9950         
        &    0.0166   & 1.0942   & 0.0175   &  1.0592        &  -         & -     \\
15      &    0.0102   & 3.7657   & 0.0108   &  3.7078        & -   & -                 
        &    0.0175   & 2.5014   & 0.0184   &  2.4285        & -    & -       \\
25      &    0.0100   & 16.3659   & 0.0106   & 15.7532        & -    & -                 
        &    0.0179   & 9.2563   & 0.0189   &  8.9927        & -    & -  \\       
\end{tabular}
\end{ruledtabular}
\end{table}
\endgroup  

Now, we are interested to investigate $E$ in the same three representative (1s$^{2}$ $^{1}$S, 1s2s $^{3}$S, 1s2p $^{3}$P) states of 
confined H$^{-}$ ion. It generally complements $S$ by showing an opposite behavior. To the best of our knowledge, $E$ for confined H$^{-}$ 
ion has never been investigated before. Therefore, in future, the present work may offer important guideline in this context. 

Next, Table~\ref{tab:table7} provides $E_{\rvec}$ and $E_{\pvec}$ for H$^{-}$ ion in ground state at the same  
$r_{c}$ values chosen in Table~\ref{tab:table5}. X-only (columns 2, 3), Wigner (columns 4, 5) and LYP (columns 6, 7) results are 
given. $E_{\rvec}$ progresses and $E_{\pvec}$ abates with growth in $r_{c}$. At $r_{c} \rightarrow 0$ region, X-only and 
correlated results in both spaces become very similar. Akin to $S$, with increase in $r_{c}$ this situation alters implying  
the participation of correlation contribution in density. 
Then $E_{\rvec}$, $E_{\pvec}$ for 1s2s $^{3}$S and 1s2p $^{3}$P states are presented in Table~\ref{tab:table8}. The 
arrangement is similar to Table~\ref{tab:table6}. $^{3}$S values are given in second to seventh columns, while
$^{3}$P results are given in last six columns. Similar to the ground state, here also for both states, the X-only and 
correlated results resemble each other at strong confinement limit. At low to moderate $r_{c}$ region, $E_{\rvec}$ values of 
1s2s $^{3}$S state lie higher than 1s2p $^{3}$P state and crossing occurs in between $r_{c}=$ 5 to 6. However, $E_{\pvec}$ 
follows a reverse pattern. 

\begin{figure}                         
\begin{minipage}[c]{0.33\textwidth}\centering
\includegraphics[scale=0.58]{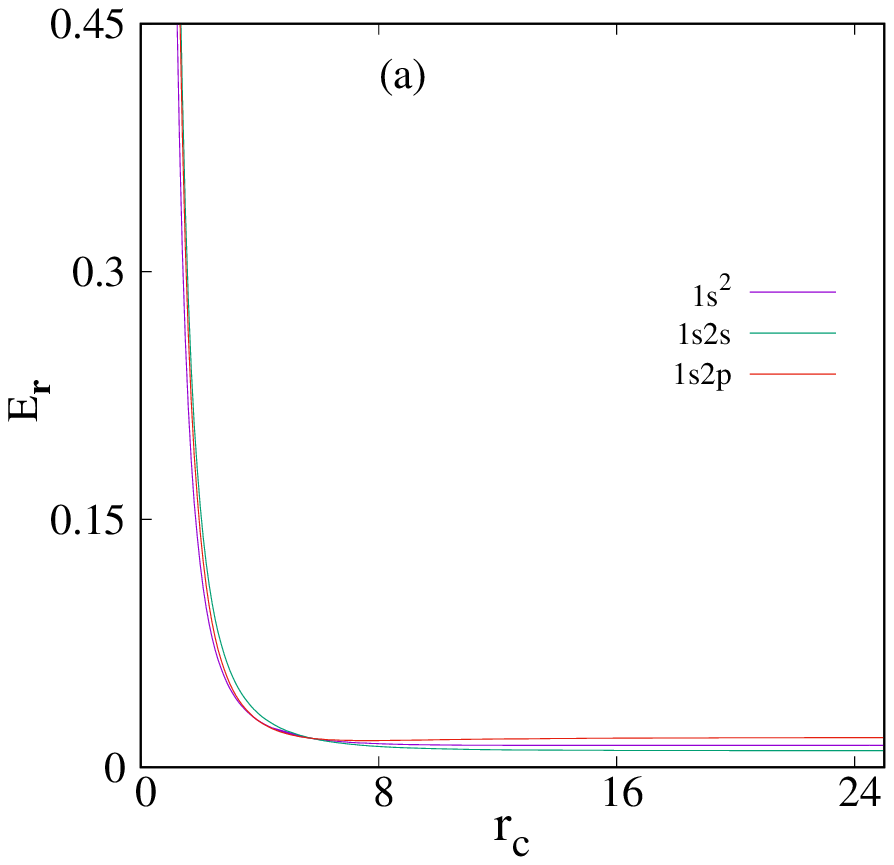}
\end{minipage}%
\begin{minipage}[c]{0.33\textwidth}\centering
\includegraphics[scale=0.58]{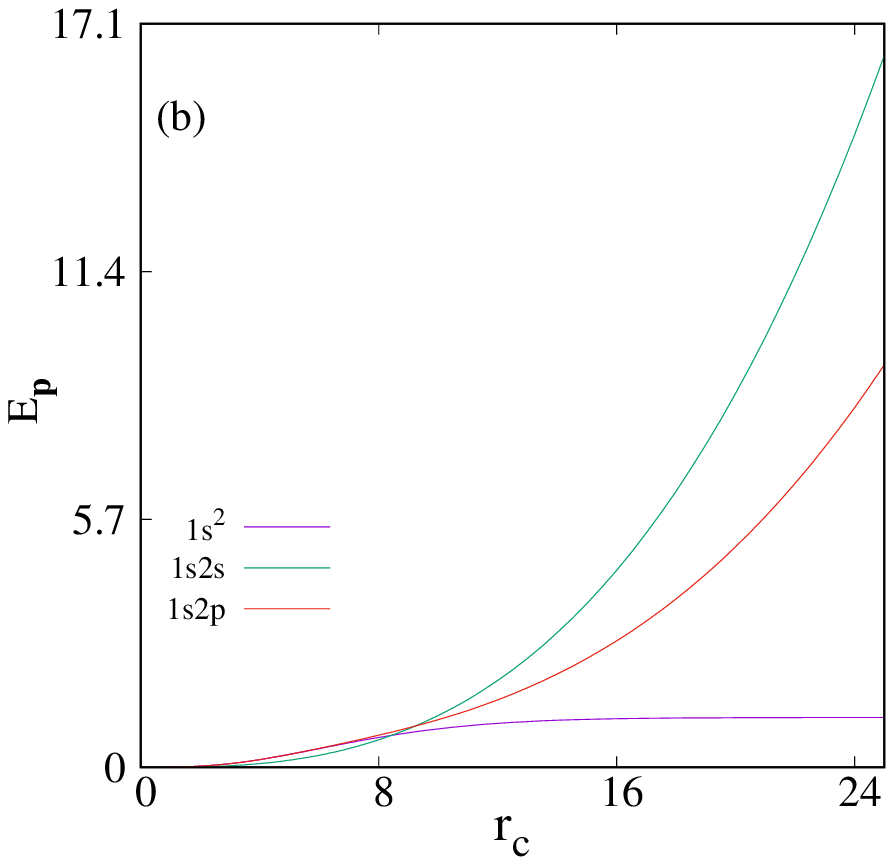}
\end{minipage}%
\begin{minipage}[c]{0.33\textwidth}\centering
\includegraphics[scale=0.58]{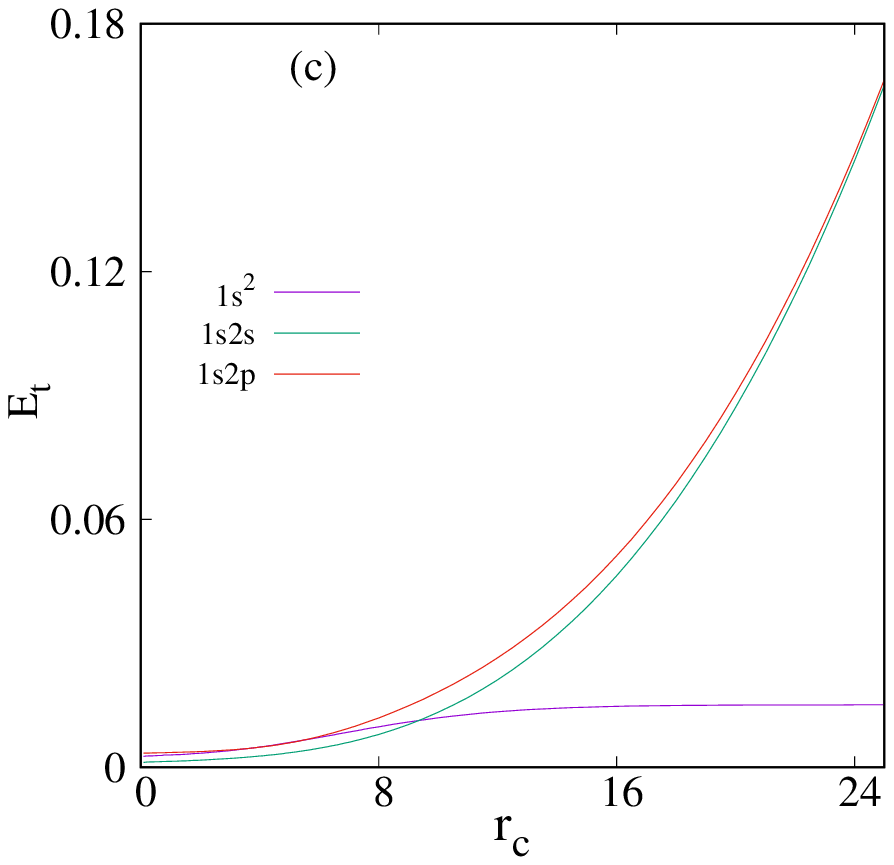}
\end{minipage}%
\caption{\label{fig:figure6} Variation of (a)$E_{\rvec} (b)E_{\pvec} (c)E_{t}$ with change in $r_{c}$ for H$^{-}$ ion.
See text for details.}
\end{figure}

Finally, $E_{\rvec}$, $E_{\pvec}$, $E_{t}$ for 1s$^{2}$ $^{1}$S, 1s2s $^{3}$S, 1s2p $^{3}$P states
are plotted as functions of $r_{c}$ in panels (a)-(c) of Fig.~\ref{fig:figure6} respectively. 
As usual here also consideration of X-only results are sufficient to illustrate the essential purpose. $E_{\rvec}$ 
declines with gain in $r_{c}$, while 
$E_{\pvec}$ accelerates. As expected, multiple crossings between states takes place but they are not prominent from panel (a).    
However, panel (b) indicates the crossover between 1s$^{2}$ $^{1}$S, 1s2s $^{3}$S and 1s$^{2}$ $^{1}$S, 1s2p $^{3}$P.
$E_{t}$ in all these three cases increase with $r_{c}$.

\section{Future and Outlook}
An appropriate and effective KS DFT method is presented for calculation of H${^{-}}$ ion trapped inside an
impenetrable spherical cavity of varying radius. The proposed recipe is computationally achievable and can easily be
applied to other atoms in both ground and excited states. Energies are reported for ground and selected singly 
excited (1s2s $^{3,1}$S, 1s2p $^{3,1}$P) states of H${^{-}}$ ion in wide range of $r_{c}$ covering strong, 
moderate and weak confinement regime.  Accurate results for a given state can be achieved, provided the exchange 
contribution is properly taken into account, which, of course, is the key reason behind the general success of this approach.
Wigner correlation energies show qualitative similar behavior with the high-quality result of Hylleraas method. 
The results are generally in good agreement with the available literature. X-only results are very close to HF. A detailed investigation 
involving the singly excited states has been done to understand the rearrangement of atomic orbitals in strong confinement region. 

In order to test the quality of the constructed density, $S, E$ in composite $r$ and $p$-spaces has been studied for 
ground and 1s2s $^{3}$S, 1s2p $^{3}$P states. To the best of our knowledge, this is the first reporting of information entropy 
in both ground and excited state of confined atoms in very strong confinement ($r_c \leq 0.1$) region. This study reinforces 
the previous conclusion \cite{majumdar20} that, at strong confinement zone, contribution of correlation effect in density is 
small. In order to increase the correctness and accuracy of the method, better correlation energy functionals are required
to be designed and incorporated. In future present method may be extended to other atoms as well. Further, it is 
encouraging to probe the current procedure for other important realistic confinement scenario (such as encapsulation of an atom in 
supramolacular cavity). Investigation of multipole polarisability, atomic avoided crossing, hyperpolarisability,    
influence of electric and magnetic field through dynamical study is highly desirable, some of which may be undertaken later.

\section{Acknowledgement}
Financial support from BRNS, India (sanction order: 58/14/03/2019-BRNS/10255) is gratefully acknowledged. SM is obliged 
to IISER-K for her Senior Research Fellowship. NM thanks CSIR, New Delhi, India, for a Senior Research Associateship 
(Pool No. 9033A).

\bibliography{ref}
\bibliographystyle{unsrt} 
\end{document}